# Did JHotDraw Respect the Law of Good Style?

## A deep dive into the nature of false positives of bad code smells

## Daniel Speicher[a]

a   Bonn-Aachen International Center for Information Technology, University of Bonn

**Abstract**   Developers need to make a constant effort to improve the quality of their code if they want to stay productive. Tools that highlight code locations that could benefit from refactoring are thus highly desirable. The most common name for such locations is "bad code smell". A number of tools offer such quality feedback and there is a substantial body of related research.

However, all these tools, including those based on Machine Learning, still produce false positives. Every single false positive shown to the developer places a cognitive burden on her and should thus be avoided. The literature discusses the choice of metric thresholds, the general subjectivity of such a judgment and the relation to conscious design choices, "design ideas".

To examine false positives and the relation between bad smells and design ideas, we designed and conducted an exploratory case study. While previous research presented a broad overview, we have chosen a narrow setting to reach for even deeper insights: The framework JHotDraw had been designed so thoughtfully that most smell warnings are expected to be false positives. Nevertheless, the "Law of Good Style", better known as the "Law of Demeter", is a rather restrictive design rule so that we still expected to find some potential bad smells, i.e. violations of this "Law".

This combination led to 1215 potential smells of which at most 42 are true positives. We found generic as well as specific design ideas that were traded for the smell. Our confidence in that decision ranged from high enough to very high. We were surprised to realize that the smell definition itself required the formulation of constructive design ideas. Finally we found some smells to be the result of the limitation of the language and one could introduce auxiliary constructive design ideas to compensate for them.

The decision whether a potential smell occurrence is actually a true positive was made very meticulously. For that purpose we took three qualities that the smell could affect negatively into account and we discussed the result of the recommended refactorings. If we were convinced that we had found a false positive, we described the relationships with design ideas.

The realization that not only general design ideas but also specific design ideas have an influence on whether a potential smell is a true positive turns the problem of false positives from a scientific problem ("What is the true definition of the smell?") to a engineering problem ("How can we incorporate design ideas into smell definitions?"). We recommend to add adaptation points to the smell definitions. Higher layers may then adapt the smell for specific contexts. After adaptation the tool may continuously provide distinct and precise quality feedback, reducing the cognitive load for the developer and preventing habituation. Furthermore, the schema for the discussion of potential smells may be used to elaborate more sets of true and false smell occurrences. Finally, it follows that smell detection based on machine learning should also take signs of design ideas into account.

# The Art, Science, and Engineering of Programming

## 1  The Challenge of Low Precision in Bad Code Smell Detection

Code quality needs the developer's attention. As the quality degrades, the effort for further evolution of the software system increases. Developers thus need to constantly improve the internal design. Tools that highlight locations where such improvement would be beneficial or even necessary are therefore highly desirable. However, if the tool raises too many false alarms, the developer might start to ignore the hints provided by the tool and get habituated to them.

Fowler and Beck compiled the most prominent list of "Bad Smells in Code" [12, pages 83–93] (a second edition was just published [11, pages 71–84]) describing "certain structures in the code that suggest (sometimes, scream for) the possibility of refactoring". Their smell descriptions are often very metaphorical ("feature envy", "shotgun surgery") and targeted at the human reader. Researchers have nevertheless operationalized these and other definitions. Lanza and Marinescu [18] presented a metric based approach. Moha, Guéhéneuc, Le Meur and Duchien [27] presented a rule based domain-specific language SADSL to specify design defects.

The field is under active research. Recently Sharma and Spinellis [36] presented a survey of 445 primary studies published between 1999 and 2016. The first open research question they mention is the high rate of false-positives and the lack of context. The literature mentions as reasons for the low precision the challenge of finding appropriate metric thresholds, the general subjectivity of such a judgments and competing design choices. Fowler and Beck [12] suggested for example that the smell "feature envy" might be a natural consequence of the "visitor" pattern. Lanza and Marinescu [18] claim that "intensive coupling" and "dispersed coupling" might be natural and less harmful for "initialization and configuration methods". Speicher and Jancke [41] suggest to take three kinds of context for smell detection into account. Speicher [38] discusses the relevance of structural context, especially the relation between the smells defined by Marinescu and Lanza [18] and the canonical design patterns [13]. The "Law of Demeter" easily gets in conflict with the method chaining aspect of "fluent interfaces" [10].

Pecorelli, Palomba, Nucci, and Lucia [31] report on a comparative evaluation of heuristic and machine learning approaches. They found that the machine learning approaches hardly deliver better precision, at least if trained on the same features that the heuristics take into account.

The most comprehensive study of the reasons for code smell false positives was presented by Fontana, Dietrich, Walter, Yamashita and Zanoni [8, 9]. Their two main categories are "Imposed [bad smells as] side effects of *conscious* design decisions" and "Inadvertent [bad smells being] created by tools that create (generate) or consume (analyze) code". Our study is motivated by the first category. We want to explore the relation between a smell definition and "design ideas". We use the latter term not only to capture conscious design choices but any concept a developer might take into consideration, when judging a potential occurrence of a bad smell.

In the following we will investigate the phenomenon of design ideas conflicting with bad smell definitions and the implications for the codification of bad smells and design ideas in an exploratory case study. A summary of this case study was already published [39].

 ## 2    Design of the Exploratory Case Study

A *case study* is "an empirical inquiry that investigates a contemporary phenomenon within its real-life context, especially when the boundaries between phenomenon and context are not clearly evident" [7, 45]. As we will see soon, the phenomena of design ideas and the conflicts with bad smells have indeed no sharp boundaries that could be defined up-front. The relevant "real-life context" in the case that we will study ranges from a few lines of code, to the specifics of the software under consideration, the frameworks used, the JDK or Java in general.

Case studies are distinguished into *confirmatory* and *exploratory case studies*. As we will argue later, the following case study has some confirmatory value, but its main intention is to explore the phenomena "design idea" and "conflict between bad smell definitions and design ideas" further. "Exploratory case studies are used as initial investigations of some phenomena to derive new hypotheses and build theories" [7]. We call a part of a program that matches a bad smell definition a *potential bad smell*. Such a match is called a *true positive*, if we would expect an experienced developer to change the program to remove the smell. It is called a *false positive*, if we expect an experienced developer to keep the current state.[1] To guide our investigation we formulate three open research questions[2] concerned with how false positives and design ideas occur and relate:

**RQ1**  What is the nature of design ideas that conflict with bad smell definitions?

**RQ2**  How can a design idea turn a potential bad smell into a false positive?

**RQ3**  What are the consequences for the definition of bad smells and design ideas?

Motivated by the goal of collecting a rich set of answers to our research questions, we devise the following *study proposition*: Since we intend to explore the phenomena, we choose a case in which we hope to find many instances of them. The choice consists of two parts. First, we choose a rather rigid bad smell definition, so that we can expect a high number of potential smells in the code. Second, we choose a thoughtfully designed piece of software, so that in case of a conflict we can be optimistic of finding a valuable design idea. We consider the single potential design smell as *unit of analysis*.

---

[1] The decision whether to change the code or not involves human judgment. To emphasize that positive and negative aspects must be weighed against each other, one may also formulate: "[A] false positive for [a bad smell] is a code structure that satisfies the conditions that define [the smell], but it is not clear that they have a net detrimental effect on the quality of the software system under analysis" [8].

[2] The role of the research question for a case study is characterized as follows [7]: "A precondition for conducting a case study is a clear research question concerned with how or why certain phenomena occur. This is used to derive a *study proposition* that states precisely what the study is intended to show, and to guide the selection of cases and the types of data to collect."

The chosen bad smell definition is the strict form of the classical "Law of Demeter"—also known as "Law of Good Style"—and will be introduced in section 3. The chosen software is "JHotDraw" in version 5.1 and will be introduced in section 4.

The crucial analysis step for each potential bad smell is the discussion of whether we consider it to be a *true positive* or a *false positive*. As we do not know upfront whether these discussions will lead to clear results, we might have to elaborate other classifications than true or false. We will describe our analysis method in 5 after the presentation of the bad smell definition and the software. Its main ingredients are the discussion of the negative impacts of the current design, the discussion of the positive consequences of the removal of the potential smell, and references to established design principles, recent empirical research about the impact of bad smells, and documented design intentions. Once we have analyzed the potential smell in this respect, we can collect the further properties targeted by our research questions. Depending on the given arguments the discussion will not only apply to the given potential bad smell but also to similar cases.

We have chosen the combined case "Law of Demeter" on "JHotDraw 5.1" with the expectation to find many instances of the phenomena under consideration. While a high number of these instances increases the value of the study as an exploratory case study the optimistic expectations decrease the study's value as a confirmatory case study. Since we are already confident of finding evidence for our work there is not much corroboration for it in actually finding such evidence.[3]

Let us now introduce the two "opponents" of our exploratory case study in the following two sections. Section 3 presents the "Law of Demeter" and section 4 the "JHotDraw" framework. After the description of the method of analysis for the potential bad smells in section 5 the actual study is presented in the section 6. The results of the case study are presented in section 7.

## 3    The "Law of Demeter" / "Law of Good Style"

The goal of our exploratory case study is to explore the conflicts between bad smell definitions and design ideas. The goal is explicitly not to evaluate the design smell definition! To get a high number of conflicts we selected with the "Law of Demeter" a rather rigid design smell definition, ignored weaker variations and did not invest in perfecting the definition upfront. As an evaluation of the "Law" it would be utterly unfair. Our exploration can also be seen as a little experiment about incremental improvements of an imperfect version of a bad smell definition. Ideally the discussion will lead us to rediscover the refinements that can be found in the literature and more. The focus of the study is nevertheless the exploration of design ideas that conflict with the bad smell definition.

---

[3] We use here a term first suggested by Karl R. Popper in 1939. He states: "Instead of discussing the 'probability' of a hypothesis we should try to assess what tests, what trials, it has withstood; that is, we should try to assess how far it has been able to prove its fitness to survive by standing up to tests. In brief, we should try to assess how far it has been 'corroborated'." [34, page 248] He states clearly that our optimistic expectations are at odds with the hope for confirmation: "Confirmations should count only if they are the result of *risky predictions*; that is to say, if, unenlightened by the theory in question, we should have expected an event which was incompatible with the theory — an event which would have refuted the theory." [33]

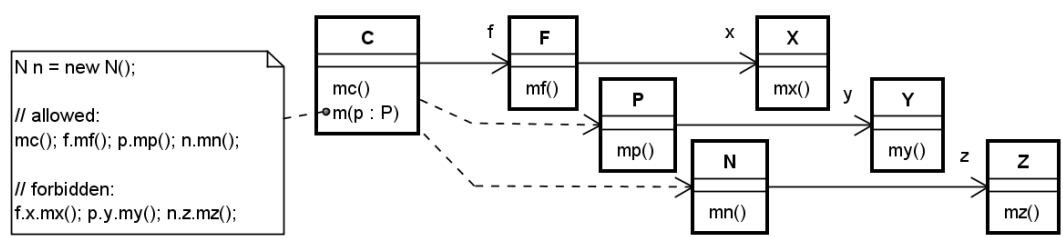

**Figure 1** Illustration of the "Law of Demeter". The classes C (class itself), F (type of a field), P (type of a parameter), N (instantiated type) are "friends" of the method m(P) in C, the classes X, Y, Z not. If the method m(P) respects the "Law of Demeter", i.e. it only accesses its "friends", it is not affected by changes in the classes X, Y, Z and a developer only needs to keep C, F, P and N in mind, when reading m(P).

Karl Lieberherr, Ian Holland and Arthur Riel had been using object-oriented programming techniques in research and teaching for two years, when they presented their answer [22] to the question "When is an object-oriented program written in good style?" They believed that their rule "helps to formalize existing ideas [about good style]" and promotes "maintainability and comprehensibility". The rule was originally suggested by Ian Holland in the context of the "Demeter" project and named "Law of Good Style" as well as "Law of Demeter".[4] The motto of the "Law of Demeter" is "Only talk to your friends" or sometimes as well "Don't talk to strangers". For every method certain types are considered "friends" and the method should only "talk" (i.e., "call a method" or "access a field") of these "friend" types. The original papers provide definitions of the "Law" for different programming languages but (of course) not for Java.

We therefore took the version for another statically typed language [21, page 47], namely C++, and translated it to Java making it even slightly more strict by leaving aside "classes of global variables", since there are no "global variables" in Java. Fields that are public and static come closest, but we postpone the decision on how to handle those until we encounter them in our case study. Method calls and field accesses are not only possible in method bodies but as well in constructors, field initialization expressions and initializer blocks. We extend the law therefore to all these *executables*.

---

[4] The Demeter project allowed the specification of a class hierarchy in a modified EBNF and to generate skeletons for applications and further utilities. The Demeter developers strove "to produce an environment which [...] allow[s] software to be 'grown' in a continuous fashion". Consequently they considered the "ease of modification [to be] one criterion which characterizes a good object-oriented programming style". The "Law of Demeter" therefore aims at limiting the impact of changes to a given program.

*Java, class form's strict version*. In all executables (methods, constructors, initializer blocks, field initialization expressions) $M$ of class $C$, you may use only members (methods and fields) of the following types (classes and interfaces) and their supertypes:

- $C$,
- types of fields of $C$,
- types of parameters of $M$, or
- classes that are instantiated in $M$.

■ **Figure 2** The Law of Demeter for Java, adapted from the C++ version.

In Java, methods and static fields can also be defined in interfaces. So, in our definition in figure 2 we address interfaces as well as classes.

Why should a developer follow the "Law"? The earlier papers listed six related principles but a later paper [20, page 71] clarifies and condenses the "Benefits of the Law" to two perspectives that we from hereon address as *limited coupling* and improved *understandability*. The "Law" limits the coupling in the sense that a method can only be affected by changes to its "friend" types. It improves understandability in the sense that "[w]hen reading a method M, the programmer has only to be aware of the [friend] classes of M". Rebecca Wirfs-Brock and Brian Wilkerson [43] suggest that *encapsulation* is the most important quality of object-oriented software and that the "Law of Demeter" somewhat[5] contributes to it. We can consider encapsulation as the degree to which an object hides its internal structure and the complexity of the implementation of its behavior. The "Law" supports encapsulation by limiting the access to classes returned by method calls that might reveal information about the internal structure of the "friend" objects. We will use these three qualities as criteria for the evaluation of potential violations of the "Law". We will have to ask ourselves whether not following the "Law" would impact encapsulation, limited coupling, understandability negatively or conversely if following the "Law" improves these qualities. Guo, Würsch, Giger and Gall [14] analyzed five Eclipse sub-projects

---

[5] The paper elaborates the idea that encapsulation can be reached much more effectively if the design process does not start from the data to be stored in the objects but from the "responsibilities" the objects have to fulfill. The responsibilities of an object are identified with the questions "What actions is this object responsible for?" and "What information does this object share?", where "[the] information shared by an object might or might not be part of the structure of that object." [43, page 73] From this perspective sharing information is not necessarily harming encapsulation, since it is not necessarily revealing the internal structure of an object. The authors consequently claim that "the Law overly constrains the possible connections between objects." Since the "Law" is suggested from a "data-driven" perspective on design, Wirfs-Brock and Wilkerson don't consider the "Law" to be as effective with respect to encapsulation as their suggestion to use Responsibility-Driven Design. They concede nevertheless that the "Law of Demeter achieves encapsulation by limiting access to objects."

**1) Violation of the "Law of Good Style" / "Law of Demeter"**

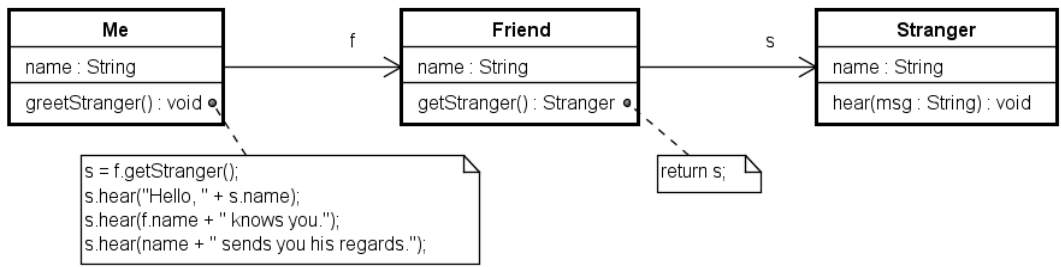

**2) Resolution: "Lifting" responsibilities "forward"**

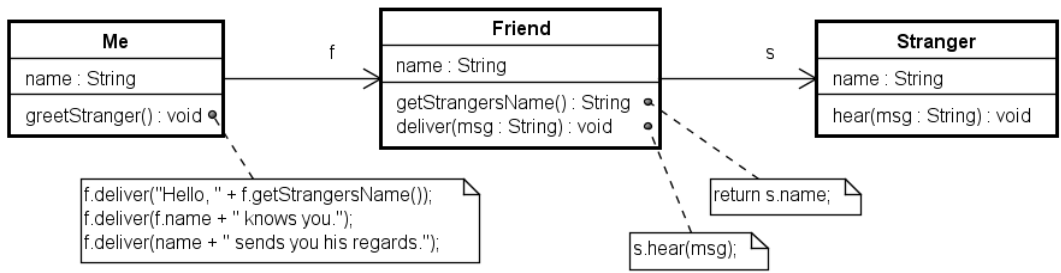

**3) Resolution: "Pushing" responsibilities "back"**

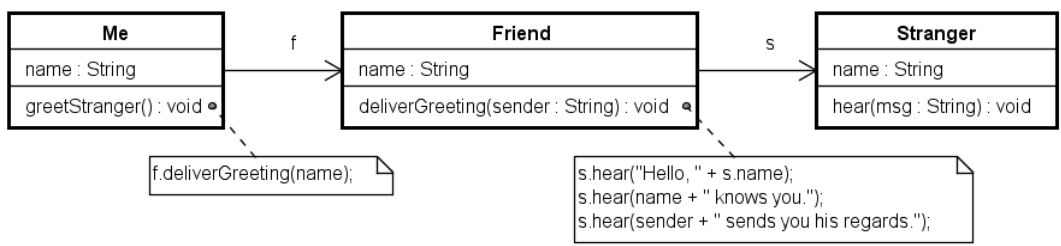

■ **Figure 3** Illustration of resolutions to "Law of Demeter" violations: In 1) the call to s.hear(String) and the access to s.name in Me.greetStranger() violate the "Law".

and found "that violations of the Law of Demeter indeed highly correlate with the number of bugs and are early predictor of the software quality".

Given a violation of the "Law", how can a developer refactor the code to remove the violation? Figure 3 illustrates two possibilities. Resolving the violation by *lifting* responsibilities *forward* means identifying the responsibilities that the violating method uses from the stranger class and letting an intermediate "friend" class take over these responsibilities. The "friend" class typically collaborates with the stranger class for that purpose. Often forwarding calls and results mechanically is already enough. As a consequence the interface of the "friend" class might grow. Resolving the violation by *pushing* responsibilities *back* means identifying meaningful parts of the original methods that rely on the stranger, extracting these into separate methods and moving them to the "friend". Again the interface of the "friend" class grows. As the example illustrates some methods in the interface of the "friend" class might well become redundant and can therefore be removed.

If the classes are part of different layers, we might use the terms *pushing down* and *lifting up*. Similar operations work as well, in the case of subclasses. We might then use the terms *pushing up* and *pulling down*.[6]

This section has introduced the bad smell definition "Law of Demeter", the qualities "encapsulation", "limited coupling", and "understandability" as criteria for the severeness of a potential bad smell and the resolutions of "pushing back" and "lifting forward".

## 4 The "JHotDraw" Project

Since our exploratory case study is meant to explore design ideas that might conflict with the given rather rigid bad smell definition, we have chosen with "JHotDraw 5.1" a thoughtfully designed piece of software, so that the chances are high that many potential smells are actually false positives and that we find a valuable design idea in its context.

JHotDraw and its predecessor HotDraw written in Smalltalk are frameworks for the creation of interactive drawing applications. They played an important role in the evolution of object-oriented programming and design. HotDraw was developed by Kent Beck and Ward Cunningham. The first ever created CRC Cards ("Class Responsibilities Collaborators" Cards, [4]) were created in an effort to document the design of the first HotDraw program [6] and are still available [5]. Ralph E. Johnson wrote the first paper that transfered the idea of design pattern from the domain of architecture [1] to software development [15] covering again HotDraw. The Java version of HotDraw now

---

[6] Abstract classes are "high" in the class hierarchy but in "low" layers. Concrete classes are "low" in the class hierarchy but "high" in layers.

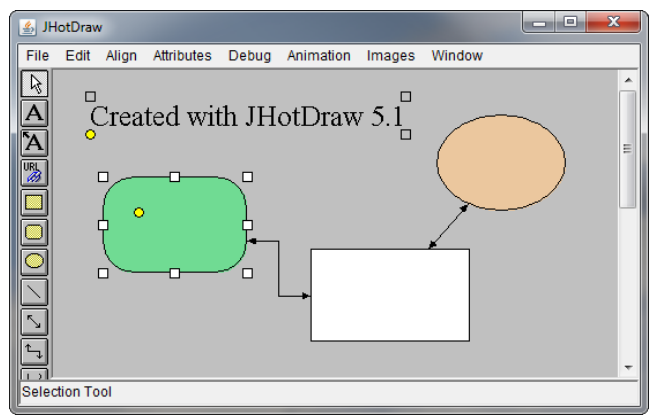

■ **Figure 4** JHotDraw 5.1 Screenshot. The screenshot shows a running `JavaDrawApp` application with menu bar on top and tool bar on the left. The `TextFigure` and the `RoundedRectangle` on the left are selected and showing their handles.

known as JHotDraw was developed by Thomas Eggenschwiler and Erich Gamma.[7] They wrote that the software "was originally planned to be a trivial case study for an applied design patterns seminar. But then it was too much fun to just stop there..."[8] The framework not only uses many design patterns but also contains thorough documentation for them in the source code and in a series of HTML documents packaged together with the framework. Because of its relevance, the pattern documentation and the open accessibility, different versions of JHotDraw are frequently used in empirical software engineering.[9] Dirk Riehle's dissertation [35] contributes additional documentation for the framework as it elaborates the collaborations in the framework with the role modeling approach developed in his work.

## 5  Method of Analysis for Potential Bad Smells

The case that we study has two factors. Both of them can be considered rather *extreme*. The "Law of Demeter" is a rather rigid classical rule, leaving no room for gradual adaptations by e.g. modifying thresholds. Accessed members belong to a class that either is a "friend" of the method under consideration or it is not. Our discussion will therefore be centered around the question, which access to which classes may be regarded as "friend access". As JHotDraw was developed not only to be used as a framework but as well polished to present object-oriented design ideas, especially design patterns, we can expect to find a rather high quality piece of software. As we elaborated above, we expect interesting findings from this combination.

In the next sections we will present and discuss the potential violations of the "Law of Demeter" found in JHotDraw. The examples will be roughly ordered by the number of violations but also grouped by similarity to allow for a better flow of the discussion. The individual potential violation is our *unit of analysis*. We will nevertheless discuss similar violations together. Our discussion will be guided by the following questions:

**A1) How bad is this potential bad smell here?**  What is the negative impact on coupling, understandability, encapsulation? Does the created coupling to the "non-friend" class increase the maintenance effort? Does the potential smell really increase the mental effort for understanding the code? To what degree is encapsulation broken, i.e. what information about the internal structure of "friend" classes or their behavior is revealed and how harmful is that? The given case might suggest further criteria.

---

[7] Dirk Riehle made the source and documentation available in the context of his dissertation at https://riehle.org/computer-science/research/dissertation/appendix-e.html.

[8] See the file documentation/drawBackground.html in the JHotDraw 5.1 distribution.

[9] Ekaterina Pek and Ralf Lämmel [32] reviewed the publications of seven major software engineering conferences in 2011/2012 with respect to their usage of corpora, i.e. "collections of software artifacts [...] to derive empirical evidence from". In the 175 reviewed papers they found 168 corpora consisting of projects. Although they found "no frequently used projects or corpora across all papers", JHotDraw has with 15 still the highest frequency of inclusion in corpora. It is only competing with eclipse. Some corpora include sub-parts of eclipse. If these sub-parts count for the category "eclipse" this category has the highest frequency, namely 22.

**A2) How good could the code be after a refactoring?** What is the positive impact of applying a "lift forward" or a "push back" refactoring? Are these refactorings possible? How good would the code be in terms of coupling, understandability, encapsulation? How many methods would we need to add to a "friend"'s interface to "lift forward"? Would the responsibility that we "push back" make sense in the target "friend class"? Can we expect it to be used by other methods? Depending on the case other refactorings beyond these two might be useful.

**A3) Is the potential bad smell a true or a false positive?** Given the answers to the previous two questions, is the code good enough or should it be refactored? Besides the given answers we might as well take comments from the literature about the "Law of Demeter" or about the impact of bad smells into account. Remarks about intentions of the code in comments and documentation might be helpful as well. The question might require other answers than just a "yes" or "no".

In case of a true positive we note the suggested refactoring, where it should be applied and how many true positives are resolved by it. Otherwise, we found a false positive and its exploration should contribute to the answers to our research questions RQ1–RQ3. Since the questions are conceptual, we postpone the deeper discussion till the end. The similarities and differences between the individual findings will help us to develop an understanding of the phenomena and our answers will be less ad-hoc, once we have collected a larger and potentially diverse set of findings. During the discussion of the individual potential smell, we restrict ourselves to more pragmatic and direct questions:

**I1) What is the design idea leading to the false positive?** What is the main idea in the code that made us judge the potential bad smell to be a false positive?

**I2) How can we improve the design smell definition?** How can we adapt the bad smell definition, to take the design idea into account and remove the false positives?

**I3) How can we characterize the extension of the design idea?** To which program elements does the design idea apply? How can we identify these program elements?

**I4) How high is our confidence in our bad smell improvement?** To what degree are the arguments for the specific finding valid for whole extension of the design idea?

**I5) How many similar cases can be found in the program?** Given that we know how to identify the extension of the design idea and how we should adapt the smell definition, we can count the cases to which the same adaptation applies.

A spreadsheet containing the names of the accessing classes and their executables, the accessed members and their classes and as well our suggested categorization is available on-line.[10]

---

■ **Table 1** Twelve methods account for a quarter of all potential violations (PV). After successively filtering out false positives based on generic and JHotDraw 5.1 specific design ideas six true positives (TP) will remain. They can be solved by lifting the access to figures through Connectors forward (D3).

| Method | PV | Generic[a] | | | | | | JHD[b] | | TP |
|---|---|---|---|---|---|---|---|---|---|---|
| TriangleFigure.polygon() | 74 | - | - | - | - | - | - | - | - | - |
| ShortestDistanceConnector.findPoint(..) | 34 | 14 | - | - | - | - | - | - | - | - |
| AlignCommand.execute() | 31 | 11 | 11 | 8 | 8 | 8 | 8 | - | - | - |
| BorderDecorator.draw(.) | 26 | - | - | - | - | - | - | - | - | - |
| ElbowHandle.constrainX(.) | 21 | 11 | 9 | 9 | 2 | 2 | 2 | 2 | 2 | 2 |
| ElbowHandle.constrainY(.) | 21 | 11 | 9 | 9 | 2 | 2 | 2 | 2 | 2 | 2 |
| ElbowConnection.updatePoints() | 21 | 13 | 10 | 2 | 2 | 2 | 2 | 2 | 2 | 2 |
| TriangleFigure.connectionInsets() | 20 | - | - | - | - | - | - | - | - | - |
| DrawApplet.setupAttributes() | 15 | 15 | 8 | 6 | 6 | 5 | 5 | - | - | - |
| DiamondFigure.polygon() | 14 | - | - | - | - | - | - | - | - | - |
| PasteCommand.execute() | 13 | 9 | 7 | 5 | 2 | 2 | 1 | 1 | - | - |
| NorthWestHandle.invokeStep(.....) | 12 | 4 | 2 | 2 | - | - | - | - | - | - |
| Sum | 302 | 88 | 56 | 41 | 22 | 21 | 20 | 7 | 6 | 6 |

[a] The generic ideas taken into account are from left to right: "Data Class" (D2), "Globally Accessible Member" (D4), "Collection Type" (D5), "Constituting Constructor Parameter" (D15), Types from java.lang (D7), "Singleton" (D10).

[b] The JHotDraw 5.1 specific ideas taken into account are: "DrawingView manages the selection" (D21), "Clipboard contains Figures" (D22).

## 6  Discussion of the Potential Violations

In JHotDraw 5.1, one can find 5858 field accesses and method calls. The application of our Law of Demeter definition in figure 2 yields 1215 or 21 % potential violations. Table 1 lists the 12 methods with the highest number of potential violations.

To demonstrate the method described in the previous section, we present in section 6.1 the first discussion (D1) in all its painstaking small steps. Such diligence is necessary for two reasons: First, our discussions present well-founded judgment rather than proven absolute truth. Second, our conceptual model is not set but only evolves through the discussion. This paper includes in the Appendix A all discussions starting from (D2) in full detail. Section 6.2 presents the discussions (D2)–(D4) still in some detail, including our first example of true positives in discussion (D3). Starting in section 6.3 from (D5) we will only give a high level overview of our findings.

### 6.1  Observations in TriangleFigure.polygon()

TriangleFigure.polygon() has 74 potential violations, that is already 6.1 % of all potential violations. So, we start our discussion with this method. Given this high number of potential violations in the method polygon(), we might expect to find a difficult to understand method that has many deep couplings to unrelated classes. Instead, as the extract of the method in table 1 shows, the method is obviously far from being difficult. It creates an empty polygon and adds three points to it depending on the rotation of

**Did JHotDraw Respect the Law of Good Style?**

■ **Listing 1** Extract from the class TriangleFigure in JHotDraw 5.1 (18 of 174 lines)

```
1   public class TriangleFigure extends RectangleFigure {
2
3     protected int fRotation = 0;
4
5     public Polygon polygon() {
6       Rectangle r = displayBox();
7       Polygon p = new Polygon();
8       switch (fRotation) {
9       case 0:
10        p.addPoint(r.x+r.width/2, r.y);
11        p.addPoint(r.x+r.width, r.y+r.height);
12        p.addPoint(r.x, r.y+r.height);
13        break;
14      // [ ... 7 more cases of this kind ... ]
15      }
16      return p;
17    }
18
19  }
```

The method polygon() has 74 potential violations, highlighted with a brown wavy underline. All of these are accesses to the fields x, y, width, height of the class Rectangle. Since they neither make the code difficult to understand nor create dangerous coupling, our discussion in (D1) leads to the suggestion to consider these 74 cases and all other accesses to the fields of the class Rectangle to be false positives.

the triangle. Because the method instantiates Polygon, the class Polygon is considered a "friend" and the calls to addPoint(int, int) are not violations. All 74 potential violations are accesses to the fields x, y, width or height of the class Rectangle.

**(D1) Accesses to Rectangle as false positives.** We discuss accesses from the class TriangleFigure to a Rectangle received from the superclass RectangleFigure. Let us discuss these potential violations of the "Law" with respect to the questions A1)–A3) to see, whether the first impression that all these potential violations are not severe withstands a systematic discussion. In the case that we decide to consider the violations to be false positives, we will discuss the findings as well with respect to the questions I1)–I5).

A1) Do these accesses create a difficulty in understanding the code? Do these accesses create additional maintenance effort, because of deep coupling? Obviously neither of the two problems are present here: The concept of a rectangle is part of our elementary common knowledge. From the usage in this method the Rectangle rather seems to be more of a record containing four data values than of a real object. If we explore the Rectangle class, we do find some methods, but all are straightforward operations on the four fields or create new rectangles. So, the usage of Rectangle is easy to understand. In addition the class is stable[11] and its structure is shallow, so that any dependency on this class does not create notable maintenance complexity. Finally we have to ask ourselves how severely the encapsulation of `RectangleFigure` is broken by the access to the fields of a `Rectangle` retrieved through its method `displayBox()`. This is again not problematic since throughout the application all figures are responsible for a rectangular area (their "display box"), so this is no secret. Whether this rectangular area is represented within the class `RectangleFigure` by an object of class `Rectangle` is still hidden, so that the decision of the representation could still be changed without affecting other classes. Summarizing, the accesses are not problematic from the perspective of coupling, understandability, and encapsulation.

A2) Could the code be improved by "lifting forward" or "pushing back"? To "lift" the access to the attributes of `Rectangle` "forward" to `RectangleFigure` we would need to replace the one method `displayBox()` by four methods, one for each attribute. This would increase the complexity of the code and it would not improve its understandability since it would replace one perfectly clear concept ("rectangular display box") by four concepts of at least the same difficulty (e.g. "height of the display box"). For the option of "pushing back", we need to identify a responsibility in the method `TriangleFigure.polygon()` that could be "pushed" to `RectangleFigure` so that no access to `Rectangle` is necessary anymore. The most promising candidate would be the creation of a polygon in the shape of a triangle. Already the class names tell us that this responsibility belongs to the `TriangleFigure` where it already is and not into the `RectangleFigure`. Furthermore, we see no other class that would use the pushed responsibility. So, neither "lifting forward" nor "pushing back" would improve the code.

A3) The coupling to `Rectangle` is not harmful because of the stability of the class, the understandability is not challenged since `Rectangle` is such a simple concept, and encapsulation is not violated here since only common knowledge is "revealed". "Lifting forward" is no option, since it just increases the size of the interface of the class `RectangleFigure` and "pushing back" is here no option, since it would misplace a responsibility. So, the answers to A1) and A2) clearly suggest that the 74 potential violation of the "Law" do not represent a bad smell and are thus false positives.

I1) Our discussion of the potential violations was dominated by considerations about the class `Rectangle` itself. So we stay for the moment with the idea "rectangle" as reference point for harmless potential violations. I2) Since our judgment relied so strongly on the properties of `Rectangle` itself, we suggest to adapt the "Law" by considering all accesses to `Rectangle` to be acceptable or, to put it more figuratively, `Rectangle` is "everybody's friend". I3) The idea "rectangle" is present in the code in the class `Rectangle` in the package java.awt. The extension of the idea "rectangle" can thus be identified by referring to the name and package of the class. I4) The observation of the harmlessness of the coupling, the understandability, and the unacceptable price for "lifting forward" just relied on the properties of the class `Rectangle`. We are optimistic

---

[11] The class is part of the package java.awt. Its simplicity makes changes improbable. The mere amount of code depending on it makes changes to it too expensive, i.e. it is also stable in the sense of Robert C. Martin [26].

■ **Listing 2**   Extract from the class `ElbowHandle` in JHotDraw 5.1 (27 of 126 lines)

```
1  public class ElbowHandle extends AbstractHandle {
2
3      private int fSegment;
4
5      public ElbowHandle(LineConnection owner, int segment) {
6          super(owner);
7          fSegment = segment;
8      }
9
10     private int constrainX(int x) {
11         LineConnection line = ownerConnection();
12         Figure startFigure = line.start().owner();
13         // [... similar code for endFigure deleted ...]
14         Rectangle start = startFigure.displayBox();
15         Insets i1 = startFigure.connectionInsets();
16         int r1x, r1width;
17         r1x = start.x + i1.left;
18         r1width = start.width - i1.left - i1.right-1;
19         if (fSegment == 0)
20             x = Geom.range(r1x, r1x + r1width, x);
21         return x;
22     }
23
24     private LineConnection ownerConnection() {
25         return (LineConnection)owner();
26     }
27
28 }
```

that as well in other cases an intermediate "friend" class returning a `Rectangle` does not break its encapsulation by doing so. The effect of "pushing back" relies more strongly on the method under consideration and the intermediate "friend" class. Nevertheless our confidence in the harmlessness of accesses to `Rectangle` is high enough to keep this general adaptation. I5) The decision to ignore all accesses to `Rectangle` already reduces the number of potential violations by 366 or 30.1 %.

## 6.2  Observations in `ElbowHandle.constrainX(int)`

The responsibility of the method `constrainX(int)` in the class `ElbowHandle` shown in listing 2 is the following: Figures may be connected by `LineConnections` consisting of horizontal and vertical segments. The user can move them with handles. The class `ElbowHandle` ensures that the first and the last segment are not moved beyond the inner boundaries of the start or the end figure respectively. The necessary calculation for this constraint is implemented in the methods `constrainX(int)` and `constrainY(int)`.

**(D2)  Accesses to "Data Classes" as false positives.**   In `constrainX(int)` are four accesses to the fields `x` and `width` of the class `Rectangle`, which we decided (D1) to ignore. The field names `left` and `right` of the class `Insets` motivate the guess that we are looking at a similar class to `Rectangle` and that we might decide similarly about the accesses.[12]

A detailed discussion shows again for the same reasons as for `Rectangle` that the coupling to `Insets` is not harmful, the understandability is not challenged, and encapsulation is not really broken. "Lifting forward" would increase the size of the interface `Figure` too much and "pushing back" would misplace a responsibility. So the 6 accesses to fields of `Insets` do not represent a bad smell and are thus false positives. We suggest to give the similar character of the two classes `Rectangle` and `Insets` the name "Data Class". A "Data Class" is a record-like class containing mainly state and almost no behavior.[13] All accesses and calls to "Data Classes" in the "Law of Demeter" should then be ignored. Besides the classes `Insets` and `Rectangle` there are a few more classes that can be considered to be "Data Classes": `Color`, `Dimension`, `FontMetrics`, `Point` and `Polygon`. All these classes are in the package `java.awt`. Enumerating these class names is an unambiguous way to define the extension of the concept "Data Class". For the given method this exception already reduces the number of violations again by 6 and we reached a reduction from 21 potential violations down to 11.[14] The overall number of potential violations can be reduced by 539 cases or 44.4 % of the original violations. The reductions per class are: `Rectangle`=366, `Point`=56, `Dimension`=36, `Color`=35, `Insets`=34, `Polygon`=7, `FontMetrics`=5.

**(D3) Refactoring: Lift access to figure forward.**    As we are now able to put aside the obviously irrelevant accesses to the two "Data Classes", let us explore whether the remaining 11 violations are meaningful. From the code we understand that the `ElbowHandle` is owned by a `LineConnection`, which has a `Figure` as start and as end. If we look closer, another type is involved: `line.start()` returns an object of type `Connector`, which seems to be owned by the start figure, as the result of a call to `owner()` is stored in the local

---

[12] The JavaDoc in `java.awt` explains the class as follows: "An `Insets` object is a representation of the borders of a container. It specifies the space that a container must leave at each of its edges. The space can be a border, a blank space, or a title."

[13] Whether or not "Data Classes" are acceptable design is often heavily discussed between developers. See for example the question "Is there any reason to use "plain old data" classes?" at http://programmers.stackexchange.com/questions/30297/ for a liberal position. In the context of JHotDraw the existence of objects containing geometrical information is very natural. Empirical research [44] suggests as well that "Data Classes" are not a reason for maintenance problems, but rather the contrary. Robert C. Martin suggests that the "Law of Demeter" should not be applied, when accessing "Data Classes" [25]. Furthermore, he recommends "Data Structures" as means to pass data across boundaries of the architecture [24]. While the first edition of Fowler's book on refactoring [12] mentioned "records" on just half a page, the second edition [11] spends almost eight pages on the same topic and has multiple mentions more. While the book still recommends to replace "records" with "Data Classes" and to later enrich these classes, data has a much more natural place in the discussion. Fontana, Dietrich, Walter, Yamashita, and Zanoni [8, IV B 2)] collected evidence that "Data Classes" have "only little negative impact on maintainability of the subject class".

[14] For the reduction in the 12 methods with the highest number of violations see table 1.

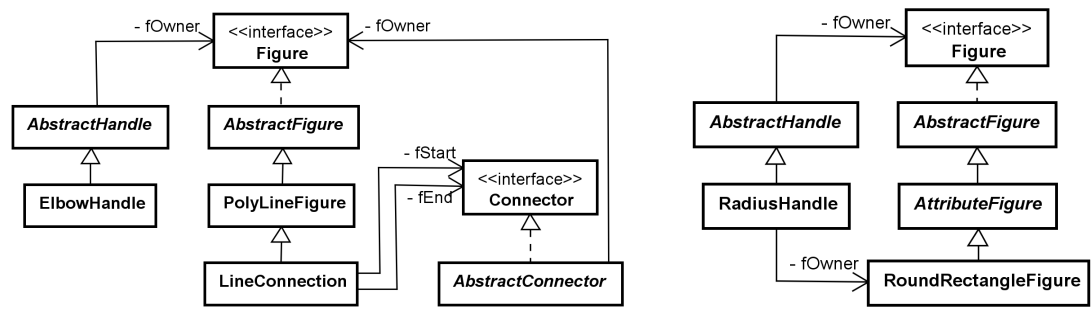

**■ Figure 5**  Class diagrams for `ElbowHandle` and `RadiusHandle`. Details in Figures 7 and 8.

variable startFigure. If we examine the call of `Connector.owner()` on the result of `line.start()` thoroughly, it becomes clear that 1) this is exactly the kind of dependency on internals of "non-friend" class the "Law of Demeter" discourages and that 2) this is indeed an unnecessary complexity hindering understanding and creating unnecessary obstacles to the further evolution of the relation between `LineConnection`, `Connector` and `Figure`. Lifting the access to `Connector.owner()` by using the already existing methods `startFigure()` and `endFigure()` resolves two true violations in `constrainX(int)`, two in `constrainY(int)` and two more in `ElbowConnection.updatePoints()`.

**(D4) "Globally Accessible Member"**  The method `constrainX(int)` contains two calls to the `public` and static method `range(int, int, int)`. Accesses to static members don't need a reference to an object so there is no natural relation to the types of fields, parameters, instantiations and no way to apply the "Law". Let us call `public` static members "Globally Accessible Member" and consider them to be "everybody's friend". This reduces the number of potential violations by 270 cases or 22.2 % of the original violations.

### 6.3  Observations in `StandardDrawingView.drawHandles(Graphics)`

The `StandardDrawingView` has a field `fSelectionHandles` containing the handles of the selected figures. The method `drawHandles(Graphics)` gets an `Enumeration` containing these handles and calls `draw(Graphics)` on each. The calls to the methods of `Enumeration` and the call to `draw(Graphics)` are potential violations. However, `Enumerations` and collection types in general are such an essential part of Java that we should consider all collection types "everybody's friend" (D5). `Handle` should be considered a "friend" of the `StandardDrawingView` as the field realizes an aggregation of `Handles`. An aggregation or any association with higher multiplicity should be considered as "befriending" as an association with multiplicity one. The aggregated type could unfortunately not be expressed up to Java 1.5 and thus needs to be inferred, guessed, or just named (D6). **Cases:** (D5) "Collection Types" 137; (D6) "Known Type in Aggregation" 40.

### 6.4  Observations in `IconKit.loadImageResource(String)`

The method `loadImageResource(String)` in the class `Iconkit` loads an image identified by the location of the class file and the method parameter. It starts by accessing a "Singleton"

Toolkit followed by four potential violations: The Class of the current instance is asked for an URL, the URL is used to retrieve the file content, the Toolkit is used to create an Image from the content. In between is a call to System.out.println(String). As we had already argued to consider "Collection Types" to be part of the language, this should hold a fortiori for classes in java.lang like Class (D7). The "Standard Printing Idiom" is not favored by every developer, yet the "Law" is not reason enough to forbid it (D8). Since an URL is rather a value than an active object, it could be considered a "Data Class" (D9). As the "Law" allows to access types instantiated in the method, "Singletons" must be allowed in methods calling its "Accessor" (D10). Similarly the use of other creational design patterns should be like instantiation (D11). **Cases:** (D7) "Types of the Java Language" 128; (D8) "Standard Printing Idiom" 11; (D9) java.net.URL as "Data Class" 1; (D10) "Singleton" Design Pattern 24; (D11) Further Creational Design Patterns 3.

### 6.5 Observations in JavaDrawApp.createWindowMenu()

The method createWindowMenu() in the class JavaDrawApp creates a menu, adds a menu item, and connects it to an anonymous inner class implementing the ActionListener interface. The only method of this anonymous inner class calls a method of the outer class JavaDrawApp. Since the inner class did not get to know the outer class in any of the four ways listed in the "Law", this call is a potential violation. When crafting our Java version of the "Law of Demeter", we did not take this language feature into consideration. Inner classes are meant to access their outer class and we suggest to consider all "friends" of the enclosing method to be as well "friends" of the "Anonymous Inner Class" (D12). Notice that this makes the definition of "friend" recursive. **Cases:** (D12) Inner classes share the friends of the outer class 23.

### 6.6 Observations in subclasses of AbstractHandle

Handles are displayed as small squares or circles on a figure and let the user manipulate the figure using the mouse. The figure owns its handles.[15] AbstractHandle realizes this association through a field fOwner of type Figure accessible to its subclasses through owner().[16] Some handles need a covariant owner type, i.e., the specific handle needs to know that its owner is a special kind of Figure. This is not easy to express without Java Generics and the code contains five different attempts to do it.[17] The expected specific type may be expressed as constructor parameter or not. The result of owner() may be downcast in a dedicated method or just where needed. Alternatively the class may keep a reference to the owner in a redundant field of the more specific type (D13). Given the close collaboration between the handles and the figures, we argue that figures should be considered "friends" of their handles. For the "Covariant Redundant Field" this is already true. In the other cases we find potential violations. These are helpful to acknowledge the disparity of the covariant owner representations. But, it would be nice to have a straightforward reason to consider the specific figures "friends" of the specific handles (D14). In all cases the constructor parameter should have the specific figure type. Most — in JHotDraw 5.1 all — constructor parameter are used to initialize fields thus we could consider them to be like fields and their types "friends" of the class (D15). Where the specific type is not used yet, this may be changed through a refactoring (D16). **Cases:** (D15) "Constituting Constructor Parameters" are like fields 86; (D16) Refactoring: Specialize Constructor Parameter Type 19.

---

[15] Handles on figures can be seen in the screenshot in figure 4.
[16] See the class diagrams in figure 5 or figure 7 and figure 8 in the Appendix.
[17] Table 2 in the Appendix gives a detailed overview.

### 6.7 Observations in DrawApplication

A `DrawApplication` constructs `Menus` and a `ScrollPane`. In `selectionChanged(DrawingView)` the calls to `checkEnabled()` on two `CommandMenus` are potential violations. The references to these two menus are retrieved from the `MenuBar` and need to be downcast. We recommend to store references to the menus as fields. This renders the cast unnecessary and makes `CommandMenu` a "friend" of the class. (D17). The expression "fonts.length", where fonts is an array of `String`, is like a field access. As this "field" does not belong to any class the "Law" would never allow access. Let us instead consider it "everybody's friend member" (D18). The method `createContents(StandardDrawingView)` contains two calls to the method `setUnitIncrement(int)` of the type `Adjustable`. The instances are retrieved from a newly instantiated `ScrollPane` via accessors `getVAdjustable()` and `getHAdjustable()`. Since the `ScrollPane` gets manipulated through calls to the `Adjustables` the encapsulation of `ScrollPane` is strongly violated. This design is in direct opposition to the "Law of Demeter". Nevertheless, the code of the JDK gives us no chance to call `setUnitIncrement(int)` other than to retrieve objects through two "Designated Accessors in External Code". So, we have to consider the result of calls to these accessors as "friends" (D19). **Cases:** (D17) Refactoring: Keep specialized references 4; (D18) "Array Length" 8; (D19) "Designated Accessor in External Code" 5.

### 6.8 Observations in SelectAreaTracker

The class `SelectAreaTracker` is responsible for drawing a rectangle ("rubber band") on the view in response to mouse actions and with the goal to finally select the figures enclosed in the rubber band. `SelectAreaTracker` contains in the method `drawXORRect(Rectangle)` three calls to methods of the class `Graphics`. The instance of `Graphics` is retrieved from a method declared in `DrawingView` which has *no implementation* in JHotDraw 5.1 but is implemented in `java.awt.Component`, the superclass of the superclass of the superclass of `StandardDrawingView`. Pushing the responsibility of the whole method `drawXORRext(Rectangle)` to `DrawingView` is a good choice. The `getGraphics()` method could then be deleted from the interface. The responsibility is consistent with the overall responsibilities of the view, namely drawing and selection management (D20). **Cases:** (D20) Refactoring: Push drawing the "rubber band" back 5.

### 6.9 Observations in `AlignCommand.execute()`

The class `AlignCommand` is responsible for aligning the currently selected figures. Its method execute() manipulates `Figures` retrieved from a `FigureEnumeration` retrieved from the `DrawingView` through selectionElements(). As `Figure` is not yet a "friend" these are potential violations. Managing the selection is one of the main responsibilities of the view, keeping the state is the responsibility of the drawing and manipulations are in commands and tools. Since "`DrawingView` Manages the Selection" of `Figures`, `Figure` should be considered "friend" for methods calling selectionElements() (D21). The `Clipboard` contains a collection of figures to be transferred. The access to `Clipboard` content is intentionally untyped, but all methods receiving the content receive `Figures` and `FigureSelections` and should thus consider them "friends" (D22). **Cases:** (D21) "`DrawingView` Manages the Selection" 20; (D22) "`Clipboard` Contains Figures" 2.

### 6.10 Observations in `ActionTool`

`AbstractTool` gives access to `DrawingView`, `Drawing`, and `DrawingEditor` through accessor methods view(), drawing(), and editor(). It receives a reference to the view from its subclasses during construction and the view gives access to drawing and editor. In the subclass `ActionTool` the editor is not a "friend". The call toolDone() on the result of editor() is thus a potential violation. It is a true violation! There is no reason why the tool should have access to the editor just for the same call. All other tools use the editor just for the same call. Extracting the call into a method and pushing it from all tools up into `AbstractTool` renders the accessor editor() unnecessary (D23). `ActionTool` similarly accesses the drawing via drawing() to call a method on the result, but this time the other tools call different methods on the drawing. Although the access breaks the encapsulation (rather of the view, less so of the `AbstractTool`), lifting would increase an intermediate interface disproportional and pushing would misplace responsibilities. It is more natural to consider drawing a "friend" everywhere where the view is a "friend". It is like `Drawing` is a "best buddy" of `DrawingView` (D24). There is something suspicious about `DrawingEditor`. The comment claims that "`DrawingEditor` is [a] mediator [and it] decouples the participants of a drawing editor." Yet, its accessor methods view(), drawing(), and tool() expose the types of the "Colleagues" (not "Participants" as in the comment, see [13, pages 273–282]) and provide access to them. They invite violations of the "Law" and we find indeed five potential violations using tool(). Resolving them requires quite some reevaluation of the design and we suggest to consider the potential violation as "false until a design review" (D25). **Cases:** (D23) Refactoring: Push notification up to superclass 6; (D24) "`DrawingView` Provides Access to its Drawing" 28; (D25) "`DrawingEditor` Exposes Colleagues" 5.

### 6.11 Observations in `PertFigure`

JHotDraw 5.1 comes with a sample application for PERT diagrams [23]. A `PertFigure` is a `CompositeFigure` built from one `TextFigure` and two `NumberTextFigures`. Three calls to methods of these specific types are potential violations. As `PertFigure` leaves the handling of its components to `CompositeFigure` they are only known as `Figure`s. If we do not want to maintain specialized references we need to share the knowledge that `TextFigure` and `NumberTextFigure` are "friends" of `PertFigure` with the smell definition (D26). `BouncingDrawing` is a subclass of `StandardDrawing` which is a subclass of `CompositeFigure`. It relies on all components being wrapped in `AnimationDecorator`s. It leaves the handling of its components again to `CompositeFigure`. The static type information was traded for the reuse of the general mechanisms of the "Composite" pattern. Without generics we are only left with the option to add the specific information that `AnimationDecorator` is a "friend" of `BouncingDrawing` to the smell definition (D27). **Cases:** (D26) "`PertFigure` is Composite of Specific Figures" 3; (D27) "`BouncingDrawing` is composite of `AnimationDecorator`s" 2.

### 6.12  Observations in `PertDependency.handleConnect(Figure, Figure)`

The method `handleConnect(Figure, Figure)` in the class `PertDependency` is invoked after the connection between two PERT figures has been established. The method relies on the parameters being of type `PertFigure`, although the declared parameter type is only `Figure` as in the method it overrides. While in conflict to substitutability, covariant dynamic parameter types are sometimes required and realized by downcasts. We may therefore consider the type of the "Downcast Parameter" a "friend" of the method (D28). Similarly the untyped content of the `Clipboard` discussed in section 6.9 is passed to two methods. The parameter types do not tell that they contain `Figure`s. The type may either be inferred by analyzing the flow or we just name the methods where we know that `Figure`s are passed on (D29). **Cases:** (D28) "Downcast Parameter" 16; (D29) "Known `Figure` Parameter" 3.

### 6.13  Observations in `FollowURLTool`

The method `mouseMove(MouseEvent, int, int)` in the class `FollowURLTool` shows an attribute "URL" of the figure under the mouse cursor. The `Figure` is found within the `Drawing` and asked for the URL via `getAttribute(String)`. This is a potential violation. It does not seem harmful. It would not be improved by a refactoring. It seems to be a false positive. Unfortunately we do not see any design idea behind this decision and thus suggest to name this case ad-hoc and consider `Figure` to be a "friend" here. A neither insightful nor satisfying yet pragmatic decision (D30). **Cases:** (D30) Postpone Two Marginal Presumably False Positives 2.

### 6.14  Observations in `ConnectionHandle.invokeStep(int, int, int, DrawingView)`

The method `invokeStep(int, int, int, DrawingView)` in `ConnectionHandle` tracks the figure under the mouse cursor and updates the end point of the connection. It accesses the "friend" `Figure` and — in a potential breach of encapsulation — `Connector`. As we see no striking refactoring opportunity, we suggest to consider `Connector` here as a "friend" "just for now" (D31). **Cases:** (D31) Postpone a Singular Possibly True Violation 1.

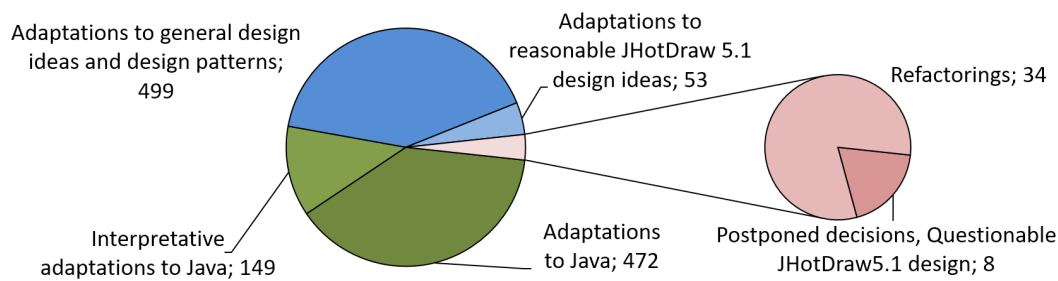

■ **Figure 6** Adaptations by numbers — Only 8 + 34 = 42 of the original 1215 cases are true positives. We thus found the high number of false positives we were hoping for.

## 7 Insights from the Exploration

Now all 1215 potential violations have been discussed in one way or another. During this discussion we encounter trivial ("length") and advanced ("inner class") language concepts as well as higher level concepts like design patterns ("Singleton"). The discussion led us to discover central design decisions ("DrawingView holds the selection"), premature solutions ("varying means to handle covariant owners"), some easy to resolve violations of the "Law of Demeter" as well as the intricate case of the "undutiful mediator".

While our main interest is in these design ideas and how they relate to the smell, we nevertheless will first have a quick look at the numbers, in section 7.1, and review the true positives and the corresponding refactorings in section 7.2. Then we answer the research questions. We start in section 7.3 with RQ2 as there are more relations between smell and ideas than just "conflict". Then we answer in section 7.4 RQ1 about the nature of the design ideas and in section 7.5 RQ3 about the consequences. Finally to comment on how our findings might or might not translate to other cases, we reflect in section 7.6 on what was special about this case.

### 7.1 The result of the discussions in numbers

For 181 cases there are multiple reasons for a potential violation to be considered a false positive. For example there are "global accessible members" in types in java.lang. The sum of the numbers from the different discussions is thus higher than the number of potential violations. We therefore group the adaptations into groups of similar generality[18] and count only those cases that are not already solved by a more general group. More general adaptations are preferable as they reduce the effort for the discussion of special cases. Figure 6 shows the numbers. The original rule had a precision of $42/1215 = 3.5\,\%$. Even without taking JHotDraw 5.1 specific ideas into account a precision of $42/(1215 - 472 - 149 - 499) = 42/95 = 44.2\,\%$ is reachable. Such a precision is nevertheless too low for continuous distinct quality feedback. Adding the specific ideas gives us at least temporarily 100 % precision and we may even silence those cases that we are not able to handle yet for a while.

---

[18] The groups for *false positives* are: Adaptations to Java (D4), (D5), (D7), (D12), (D18), interpretative adaptations to Java (D6), (D8), (D15), (D19), (D28), adaptations to general design ideas and design patterns (D2), (D10), (D11), and adaptations to reasonable JHotDraw 5.1 design ideas (D21), (D22), (D24), (D26), (D27), (D29). (In a previous publication [39] this last group accidentally did not include (D24) so that we attributed only 25 and not 53 cases to this group.) Finally there is for *true positives* a group for postponed decisions and questionable JHotDraw 5.1 design (D25), (D30), (D31) and the refactorings (D3), (D16), (D17), (D23), (D20).

## 7.2 True positives and resolving or clarifying refactorings

This study was conducted to explore the reasons for false positives. Nevertheless there would be no point in improving a smell definition that has no true findings. And indeed, after shoving the false positives to the side there are a few but substantial findings that were hidden under the pile of false positives:

**Resolving Refactorings** In three cases the refactorings introduced in section 3 were able to resolve the violations, namely in (D3) Refactoring: Lift access to figure forward, (D23) Refactoring: Push notification up to superclass and (D20) Refactoring: Push drawing the "rubber band" back.

**Clarifying Refactorings** In further cases type information was missing to recognize that an access does not violate the "Law of Demeter". The more special type would have anyway be instructive (D13), (D14) and could easily be added (D16), (D17).

**True but nasty positives** Finally we found true positives that required not only a refactoring but a local redesign (D25) `DrawingEditor` Exposes Colleagues". The comments expressed the intention to realize the mediator pattern but the code violated it.

The "Law of Demeter" has thus proven to be useful to find improvement opportunities, either directly or as a side-effect hinting on missing type information. The low original precision means there is a wealth of false positives to scrutinize. We will now summarize what we saw with respect to our research questions.

## 7.3 About the relation between the bad smell and design ideas

To answer the research question RQ2 "How can a design idea turn a potential bad smell into a false positive?" we characterize the different relations between smells and ideas that we encountered. Surprisingly "conflict" was only one of them:

**The smell definition was incomplete.** The original definition of the "Law of Demeter" did not account for inner classes (D12) and the "field `length` in arrays (D18). We deliberately left the handling of globally accessible members to our first encounter with them (D4). Furthermore we realized that types in java.lang (D7) and collections (D5) are ingredients of the language.

**The smell definition needs interpretation.** Surprisingly the apparently analytic criteria of being instantiated or being a field required interpretation. If we would consider creational design patterns (D10), (D11) different from direct object instantiation, the "Law" would forbid these means of instantiation. If a field realizes an aggregation not only the type used to hold many objects but as well the aggregated type should be considered "friend". Otherwise associations with multiplicity one would be treated differently from associations with multiplicity many (D6). Finally, a constructor parameter is expected to be stored in a field (D15) and could thus be considered to be like a field.

**The smell intention can be supported by adding type information.** If a constructor parameter is like a field, we may use its potentially more specific type (D15). Otherwise, we might rely on the return type of a "Covariant Accessor Method" (D13). We may infer the type in an aggregation from accesses (D6) and the type of a method parameter from a downcast (D28). We may furthermore manually add type information for a few cases (D22), (D26), (D27), (D29).

**Developers had chosen designs conflicting with the smell.** We found indeed cases of presumably conscious design decisions that conflict with the "Law", like sharing data through "Data Classes" (D2), (D9) or the DrawingView holding the selection and thus giving access to Figures (D21) and furthermore granting access to the model of the application, the Drawing (D24).

**Some code is beyond the control of the developer.** There is no point in worrying about cases that can not be changed as established conventions (D8) or external APIs (D19) that conflict with the "Law".

**Categorize adjourned cases for later resolution.** Finally, it may be useful to group cases that are either too hard (D25) or too unclear and irrelevant to handle them immediately (D30), (D31). Adapting the smell definition to ignore such smells for a while helps focusing on the cases that are relevant and can be solved.

The decision that a potential violation is actually a false positive relied on some factors. For the given context we were confident enough about them, but we should restate them explicitly. The coupling became less relevant for stable elements. Whether it makes sense to push some responsibility into a "friend" class depends on whether such a responsibility fits nicely in that class and whether it would be used by some clients. For JHotDraw we did not see such an opportunity, but this needs to be judged for any new system under consideration. There is as well an element of taste in these decisions. Some developers consider using System.out.print*(..) bad style and would argue against (D8). Finally the code does not necessarily contain enough information for heuristics to capture the type in an aggregation (D6).

### 7.4 About the nature of the discovered design ideas

To answer the research question RQ1 "What is the nature of design ideas *that conflict with* bad smell definitions?" let us summarize all design ideas here. Those that are actually in conflict are marked with an asterisks (*) but we have seen that there is a range of other relations:

**Language and JDK elements** Inner classes (D12) and the length of an array (D18) are part of the language. The types in the package java.lang (D7) and collections (D5) may also be seen as part of the language. The combination of the modifiers public and static characterize (D4) "Globally Accessible Member".

**Idioms and design patterns**  Associations with multiplicity bigger than one are realized via arrays, enumerations and collections (D6). "Data class" is a name for a record-like type (D2)*, (D9)*. We saw as well creational design patterns (D10), (D11) and an incorrect "Mediator" pattern (D25)*. A correct "Mediator" would not conflict with the "Law". There are the two strategies to make covariant type expectations explicit, namely "Covariant Accessor Method" and "Covariant Redundant Field" (D13). Some concepts seemed almost just like language elements, but we added a little bit more of semantics: We did not consider any downcast, but only downcasts on parameters (D28). We expressed our expectation, that collections (D5) should not do much more than just managing their elements. Finally, we expressed our expectation that constructor parameters become part of the object (D15).

**Constraints by convention or an external API**  The implementation is constrained by the conventional way of printing text (D8)* and the only possibility to access `Adjustables` in AWT (D19)*.

**Design decisions in JHotDraw 5.1**  The decision to let `DrawingView` provide access to the selected `Figures` (D21)* and to the `Drawing` (D24)* seemed reasonable, while the alleged "Mediator" (D25)* should not provide accesses to the "Colleagues", especially not to `Tools`.

**Auxiliary type information**  In the absence of generic types, we had to search for type information available in the vicinity of a possible violation (D6), (D13), (D15), (D28) and even listed some JHotDraw 5.1 specific types explicitly (D22), (D26), (D27), (D29).

**Ad-hoc categorization**  We categorized three potential violations as false but marginal (D30) or as probably true and singular (D31). These are not design ideas in the code but categories we ascribe ad-hoc to selected program elements.

The listed design ideas are on different levels of abstraction ranging from language elements to JHotDraw 5.1 specific types. The signs in the code indicating an idea are also of different generality. We can not say that any "Singleton" can be identified by having a static method called "instance()" or "getInstance()". Within a project or a team such a convention makes however sense. The description of the language and JDK elements may be read as analytic definitions. A collection could be any type implementing a collection interface. But, if the type accumulates more responsibilities than just managing its elements, our arguments are not valid anymore. The collections in the JDK leave it at this narrow responsibility.

## 7.5  Consequences for the definition of bad smells and design ideas

Based on our answers to the two previous research questions we may now approach research question RQ3 "What are the consequences for the definition of bad smells and design ideas?"

With the goal of continuous distinct quality feedback in mind we explored the possibility of achieving perfect precision by taking design ideas into account when evaluating a bad smell. The design ideas that we discovered range from very generic ("anonymous inner class") to very specific ("`DrawingView` holds `Figure` selection"), i.e.

they are only present in the given code base under consideration. Our confidence with respect to the validity of an adaptation ranges from very high ("creation method access is like instantiation") to high enough ("data classes are everybody's friend"). The confidence may be impacted by how generalizable our discussion seems. It may as well include aspects of taste or consensus within a developer community, an enterprise or a team. For analytic concepts like "Globally Accessible Member" there is an analytic definition in terms for the programming language. This definition can be used to localize all program elements that fall under this concept. For higher level concepts like design patterns there is room for interpretation. For example the constructors of a "Singleton" should typically be private, but Iconkit (D10) has a public constructor and is instantiated in the application classes. We therefore suggest to rely on the expressed intentions of the programmers to locate the design ideas. While there is no general agreement about such signs, within the context of an enterprise or a code base, such an agreement may be achievable.

We thus have concepts, adaptations and localization strategies of different generality. One would like to reuse what is general and add in a specific context, what is special. The engineering solution for this is a layered presentation. In the given case one could e.g. create three layers: 1) The original definition of the "Law of Demeter", all adaptations to Java and the adaptation to generic design ideas could be placed in a lowest layer. 2) In the layer above there could be the adaptations that are specific to JHotDraw. 3) In the highest layer there might be the temporary adaptations. Notice as well, that adaptations based on design ideas can be placed in higher layers than the design ideas, e.g. "data class" is a rather generic design idea while the decision to consider "data classes" as "everybody's friend" is a bit controversial and should therefore be placed on a higher layer.

### 7.6  What was special about the case?

The "Law of Demeter" defined access restrictions based on types. Therefore all design ideas that we discovered were as well related to access rules and types. If we explore a smell that is more directed at cohesion or more at complexity, we may encounter different design ideas. The "Law of Demeter" seemed to be a plain analytic definition. That is untypical as most other smell definitions use more flowery terms. It was all the more surprising to see that concepts like "object instantiation" and "field" require interpretation.

We traveled back in time. When the "Law of Demeter" was formalized, developers defined objects that are collections of elements together with additional functionality. At times of JHotDraw 5.1 this was about to change and collections were used just to hold some objects, but there were no generic types yet in Java. With generic types all our adaptations that artificially provide type information will become unnecessary. But new questions will arise: How should the "Law of Demeter" be interpreted for parameterized types and for type variables? Another change would be dependency injection making many creational design patterns obsolete.

## 7.7 Did JHotDraw respect the Law of Good Style?

How should we answer the question stated in the title of this article? — Most adaptations elaborated (the "spirit of") the "Law of Demeter". The access to data classes (D2) was already admitted in the original articles. The access to the aggregated type (D6) might have been as well tolerated since the role of collection changed.

The "disrespect" to the "Law" just comprises what can easily be solved by the three refactorings (D3), (D23) and (D20), the reasonable cases of (D21) "DrawingView Manages the Selection" and of (D24) "DrawingView Provides Access to its Drawing" and the bold "undutiful mediator" (D25). — JHotDraw 5.1 can thus mostly be reconciled with the "Law of Demeter", just the "undutiful mediator" should become a true "mediator".

## 8 Related Work

The research literature in bad smells has been growing at least since 2002 [37]. There are even multiple surveys and systematic literature reviews. As we mentioned in section 1 Sharma and Spinellis [36] presented a survey of 445 primary studies published between the 1999 and 2016. Sobrinho, De Lucia, and Maia reviewed [37] 351 research papers published between 1990 and 2017. Alkharabsheh, Crespo, Manso, and Taboada reviewed [2] 395 research papers published between 2000 an 2017. We relied on the Sharma and Spinellis survey as our key witness for the relevance of the question of high rates of false positives. A more thorough discussion of the work on false positives would be very desirable, especially with respect to those publications made the effort to construct manually validated datasets [3, 28, 29].

The idea that what is bad design in general might be acceptable and maybe even desirable in certain contexts is not new. Kapser elaborated this even for the "number one in the stink parade", the smell "Duplicated Code" [16]. We found in this case study again that a bad smell might be the side effect of a conscious design decision, as suggested earlier [8, 38]. Fontana, Dietrich, Walter, Yamashita and Zanoni [8, 9] conducted a *meta-synthesis* based on the empirical studies conducted in 10 years before their study, theoretical examples from the grey literature, case studies from industry and open source projects. They strove for a broad coverage. The current study strove for depth. Fontana et al. suggested as categories for anti-pattern and smells that are a side effect of conscious design decisions ("imposed anti-patterns and smells"): 1.1) "imposed by Design-Pattern", 1.2) "imposed by Programming Language used", 1.3) "imposed by frameworks", and 1.4) "imposed by optimization" and two further categories that we would rather subsume under a separate category as they describe the process that led to the smells: 1.5) "imposed by porting code from a non-object-oriented programming language" and 1.6) "inherited from legacy code". "test code" is mentioned as example in a category 2.3) "caused by analysis scope". We would rather suggest to subsume the "design idea" "test code" under 1.1) or 1.3) or a new sub-category of the category of conscious choices. 1.1) and 1.3) are as well the sub-categories that overlap with our case study.

To the best of our knowledge nobody discussed the potential incompleteness of a smell definition, the need to interpret even apparently analytic concepts, and the option to add supporting [type] information.

## 9  Summary of the Contributions

This study set out to explore the nature of bad smell false positives. We reviewed 1215 potential violations of the "Law of Demeter" in JHotDraw 5.1. As described in section 5 we evaluated these potential violations with respect to the negative impact on selected qualities and with respect to the positive impact of selected potential refactorings. To transfer this evaluation schema to other smells one needs to identify the appropriate qualities and refactorings for those smells.

As described in the literature software developers sometimes make specific design choices that are in conflict with general rules such as bad smells. We used the term "design idea" to capture not only conscious design choices but any concept a developer might take into consideration, when judging a potential occurrence of a bad smell. We explored the design ideas we encountered during our review of the false positives, what relations we observe between design ideas and smells, and what the consequences for the definition of bad smells and design ideas are.

In section 7.4 we categorized the design ideas we found as "language and JDK elements", "idioms and design patterns", "constraints by convention or an external API", "design decisions in JHotDraw 5.1", "auxiliary type information" and "ad-hoc categorization". As summarized in section 7.3 we found three kinds of conflicts, i.e. situations where the smell is present but we would not expect the developer to change the code: Developers had indeed presumably consciously chosen designs conflicting with the smell. Other design ideas mark code that is beyond the control of the developer. In very few cases we suggested to make up ad-hoc categories to mark adjourned cases for later resolution. We found three further kinds of relationships that are rather elaborations of the smell definition than conflicts: We needed to extend the smell definition to cover all language elements. Some apparently analytic criteria required interpretation and we suggested auxiliary design ideas adding type information.

The two latter relations could be described as reinterpreting the terms of the smell in the presence of the code and as reinterpreting the code in terms of the smell. Similar questions may arise for other smells. The smell "Feature Envy" requires an interpretation of what counts as "access" of a "feature" and what counts as "own feature". Coupling smells might take message sending into account that does not show as static dependency. As our study exemplifies, such need for interpretation may not be obvious up-front and we would need to rely here as so often on Iterative and Incremental Development [19].

The most consequential finding of our study might be the *significance of specific knowledge*: Whether a potential violation is a true positive does not only depend on generic design ideas but as well on *specific design ideas*. Even for generic design ideas we might need to rely on *specific signs* to identify their location. Finally, there remains a residual impact of the *specific taste* of the development team.

A smell detection tool continuously providing distinct quality feedback, needs to be precise to reduce the cognitive load for the developer and to prevent habituation to false positives. Thus, to integrate also the specific knowledge we recommend adding adaptation points to the smell definitions. Higher layers may then adapt the smell for specific contexts.

Our case study seems to indicate that precision in smell detection is only available at the cost of generality, and generality only at the cost of precision. In a layered approach higher precision may be realized by additional adapting layers and higher generality by leaving higher layers aside.

Giving up generality for precision has its own *scientific challenges*. What meaning does it have to evaluate the impact of a certain smell if its general definition is not precise? How should we compare tools if we know that general smell definitions have low precision and the differences in the tool might be just mainly on the subset of the false positives?

Developing operational bad smell definitions together with adaptations and design ideas is an *engineering task*. We have sketched how to integrate the evolution of these definitions with the evolution of the main software in a short publication [40]. Actually conducting this engineering process for our given case would have delivered a precise "Law of Demeter" definition with adaptation points and adaptations, substantial insight into specific design ideas in JHotDraw 5.1 and relevant refactoring suggestions.

**Acknowledgements**   The author likes to thank first and foremost Prof. Dr. Armin B. Cremers for his support over a long time — especially for the thorough discussions while the author was on a deep dive into JHotDraw 5.1 in an effort to sense the design considerations present in the code. While the paper now elaborates a well described way to its conclusions ($\mu\acute{\varepsilon}\vartheta o\delta o\varsigma$) the exploration started with much less orientation. Repeatedly revisiting the premature findings required strong perseverance not only from the author but probably even more from his advisor. Michael Mahlberg offered a critical second opinion on the evaluation of the object-oriented structures in the code. Günter Kniesel, Jörg Zimmermann, and Pascal Welke read an earlier version of this paper and provided valuable feedback. The anonymous reviewers provided very substantial and helpful suggestions. Not all of them could already be considered appropriately in this paper. One of the reviewers helped to reduce the number of sentences communicating that there is a German writing in English. The authors of the "Law of Demeter" papers and the developers of JHotDraw created their work with much care. This is palpable in the artifacts and made a deep examination worthwhile. Although the first paper on the "Law of Demeter" claimed to formulate an "*objective sense of style*", the truth we sensed depended substantial on specifics of the given software system. Therefore finally a thanks goes to the reader, since following into these specifics probably requires an unusual way of thinking and high reading effort.

## A  Full Discussions of the Potential Violations

In section 5 we elaborated a method for the evaluation of potential violations and demonstrated it in the discussion (D1) of the potential violations in `TriangleFigure.polygon()`. This Appendix contains all unabridged discussions starting from (D2) as a foundation for the high level overviews presented in section 6.

The author of this work made some effort to present source code excerpts and UML class diagram excerpts so that the reader can follow and evaluate the discussions without the need to consult the original source code.[19] Code excerpts are used to discuss a variety of cases, whenever possible. Although the amount of code presented is high, the reader will still need to rely on the author's report about the code every now and then. A spreadsheet containing the names of the accessing classes and their executables, the accessed members and their classes and as well our suggested categorization is available on-line.[20]

### A.1 Observations in TriangleFigure.polygon()

We dedicated in section 6.1 two pages to our discussion (D1) addressing the 74 potential violations in TriangleFigure.polygon(). All of them are accesses to the fields of a Rectangle to calculate the coordinates of a polygon representing a triangle. As these harmless accesses to fields of a simple, stable, and external class would only be avoidable if we replaced the method returning a rectangle with one method for each of its fields, we suggest to allow accesses to fields of the class java.awt.Rectangle everywhere. In terms of the "Law": We consider Rectangle to be "everybody's friend". This is the single most impactful decision as it removes 366 or 28.7 % potential violations.

### A.2 Observations in ElbowHandle.constrainX(int)

To keep the main part of this paper at a reasonable size we did not present our discussions for ElbowHandle.constrainX(int) in section 6.2 to the full extend proposed by our method described in section 5. The full discussions follow now here.

We with a method that ranks fifth with respect to the number of violations.[21] The method ElbowHandle.constrainX(int) as presented in table 3 has 21 violations. Although this method again does not look very difficult, all but the method call to the own method ownerConnection() and all but the two field accesses to the own field fSegment are potential violations. Given this imbalance between the apparent simplicity of the method and the harsh criticism by the heuristic the value of the heuristic seems for a

---

[19] The goal was to provide extracts that are as short as possible but are still syntactically correct Java code and referentially complete with respect to references within the same class, i.e., if a field or a method is referenced in a presented statement, the field or the method should have been kept as well. Fields and methods may be omitted as a whole. If some lines within a method were deleted this was indicated by an ellipsis in an end of line comment (like: // [ ... summary of what is left out ... ]]). If some tokens within a line were deleted, this was indicated by a closed comment (like: /* [...] */). Referenced methods and fields in other types as well as the other types themselves were not included. In few cases we traded the goal of syntactical correctness for the conciseness of the presentation.

[20] https://sewiki.iai.uni-bonn.de/private/daniel/public/prog2020study

[21] See table 1. ElbowHandle.constrainX(int) is the first promising candidate for further analysis: The method findPoint in ShortestDistanceConnector has 99 lines and in the methods execute in AlignCommand and draw in BorderDecorator two thirds of the potential violations, namely 20 of 31 or 16 of 24 respectively, are accesses to fields of the class Rectangle.

**Did JHotDraw Respect the Law of Good Style?**

■ **Listing 3**  Extract from the class ElbowHandle in JHotDraw 5.1 (35 of 126 lines)

```java
1   public class ElbowHandle extends AbstractHandle {
2       private int fSegment;
3
4       public ElbowHandle(LineConnection owner, int segment) {
5           super(owner);
6           fSegment = segment;
7       }
8
9       private int constrainX(int x) {
10          LineConnection line = ownerConnection();
11          Figure startFigure = line.start().owner();
12          Figure endFigure = line.end().owner();
13          Rectangle start = startFigure.displayBox();
14          Rectangle end = endFigure.displayBox();
15          Insets i1 = startFigure.connectionInsets();
16          Insets i2 = endFigure.connectionInsets();
17
18          int r1x, r1width, r2x, r2width;
19          r1x = start.x + i1.left;
20          r1width = start.width - i1.left - i1.right-1;
21
22          r2x = end.x + i2.left;
23          r2width = end.width - i2.left - i2.right-1;
24
25          if (fSegment == 0)
26              x = Geom.range(r1x, r1x + r1width, x);
27          if (fSegment == line.pointCount()-2)
28              x = Geom.range(r2x, r2x + r2width, x);
29          return x;
30      }
31
32      private LineConnection ownerConnection() {
33          return (LineConnection)owner();
34      }
35  }
```

The method constrainX(int) has 21 violations. According to the discussion in (D1) the four accesses to fields of the class Rectangle should already be considered to be false positives and are thus highlighted with a green wavy underline. The case for the six accesses to the fields left and right of the class Insets is obviously similar and the discussion in (D2) suggest to consider all accesses to fields in "Data Classes" to be false positives as well. The method ownerConnection() and the constructor parameter owner will be set into context in section A.6 in table 2 and especially in the discussion (D15). As a result of this discussion we will consider the class LineConnection and consequently its supertype Figure as "friend" of the method. The calls to start(), end(), displayBox(), connectionInsets() and pointCount() will then not be considered violations anymore. The calls to Connector.owner() on the result of line.start() and line.end() however are—as discussed in (D3)—true violations, establish an unnecessary dependency to the class Connector and can easily be removed by directly using LineConnection.startFigure() and endFigure() on line instead. These two true violations are highlighted with a red wavy underline. The relationships between the types are shown in the partial class diagram in figure 7. Finally the two unproblematic calls to range(int, int, int) will be discussed in (D4).

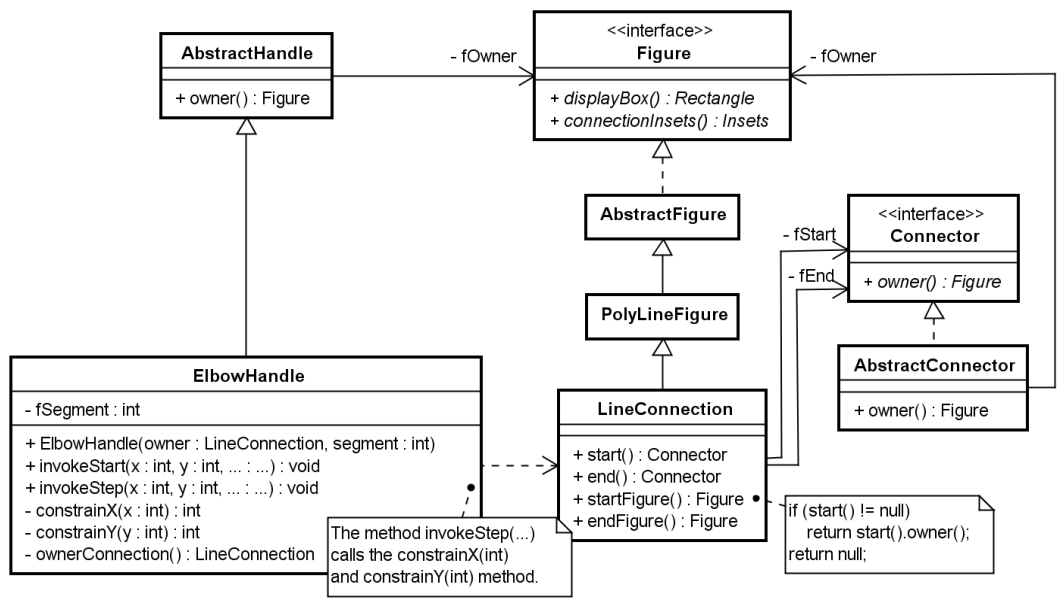

■ **Figure 7** Partial class diagram of ElbowHandle and related classes in JHotDraw 5.1—For the source code of the method constrainX(int) in the class ElbowHandle see table 3. The method startFigure() can be used to keep client methods unaware of the indirection through Connectors, so that LineConnection might get changed without the need to change e. g. constrainX(int).

moment to be at least questionable. We will have to discuss the potential violations carefully.

The responsibility of the method constrainX(int) in the class ElbowHandle shown in table 3 is the following: Figures may be connected by LineConnections consisting of horizontal and vertical segments. The user can move them with handles. The class ElbowHandle ensures that the first and the last segment is not moved beyond the inner boundaries of the start or the end figure respectively. The necessary calculation for this constraint is implemented in the methods constrainX(int) and constrainY(int).

Section A.6 will provide an extensive discussion of the relation between handles and the figures that own them. Intuitively it should be clear that ElbowHandle and LineConnection are closely related. Nevertheless the definition of the "Law of Demeter" does not immediately suggest to consider LineConnection to be a "friend" of the methods in ElbowHandle since it is never instantiated or used as a parameter or as a field in this class. Anticipating the result of that discussion we consider the class LineConnection as "friend" of the method constrainX(int). The calls to start(), end() and pointCount() are then no violations anymore. With LineConnection its supertype Figure is a "friend". Therefore the calls to displayBox() and connectionInsets() are as well no violations. (D2) discusses the accesses to the fields left and right of the class Insets. (D3) discusses the calls to Connector.owner() on the result of line.start() and line.end(). The partial class diagram in figure 7 adds some context to this discussion. Finally, (D4) discusses the calls to Geom.range(int, int, int).

**(D2) Accesses to "Data Classes" as false positives.** In constrainX(int) are four accesses to the fields x and width of the class Rectangle, which we decided (D1) to ignore. The field names left and right of the class Insets motivate the guess that we are looking at a class similar to Rectangle and that we might decide similarly about the accesses. So, we continue our methodical discussion to see, whether it actually leads us to this conclusion.

We discuss accesses from ElbowHandle to an object of class Insets received from a Figure. A1) How bad are the accesses i1.left, i1.right, i2.left, and i2.right? A quick look at the class Insets shows that it mainly consists of the four fields top, left, bottom and right and the comments explain, that these fields represent "the space that a container must leave at each of its edges". Although the idea of this class is not necessarily common knowledge, it is easy to understand once you think for a moment about how one could describe where to place figures in container figures. I.e., we are looking at a concept that is not common knowledge but a concept that is natural for the domain of layouting widgets and figures. Similar to Rectangle the class is in the package java.awt and thus stable, and the structure of the class is shallow, so that coupling to this class is no problem. The two objects of type Insets are the result of calls to the method connectionInsets(), but again there is no need to assume that the class Figure is revealing knowledge about the internal representation of its state. It is just reporting about properties that other objects might indeed need to know. Summarizing, the accesses to Insets are again not problematic from the perspective of understandability, coupling, and encapsulation.

A2) Lifting the access to the four values forward is again no option, since it would replace one method by four methods[22] of similar difficulty. A responsibility that we could push back into Figure is not easy to identify. It could be the whole calculation of acceptable x values. Would this responsibility be well placed in the class Figure? Would this responsibility be used by other classes? Unfortunately this case is much less clear than the case before and requires a deeper discussion:

The four lines in table 3 calculating r1x, r1width, r2x and r2width are essentially repeated in the method findPoint(ConnectionFigure, boolean) in the class ShortestDistanceConnector. So, is there a responsibility that should be in Figure and be used by ElbowHandle and ShortestDistanceConnector? After examining both methods closely, we find that both methods make use of the knowledge about the display box and the insets in different way. ElbowHandle constraints the acceptable movement of the last or first element of a line connection consisting of segments that are either horizontal or vertical. The class ShortestDistanceConnector calculates end points for connections of shortest length between two figures. Besides the four lines reducing the display box rectangle by the insets the commonalities of the two calculations are not obvious. If we study the calculations very thoroughly, we can find that we can express half of the findPoint(ConnectionFigure, boolean) method by the calculation in the constrainX(int) and

---

[22] For the given four violations two methods would be enough, but constrainY(int) would need the other two.

constrainY(int) methods[23] but is it true that finding the shortest distance connection should be formulated in terms of constraining the movement of line connections or that both calculations should be based on a third calculation implemented in Figure? Let us discuss this question in the spirit of Parnas [30] and ask ourselves, what kind of changes we want to encapsulate in the different classes. Should a change in one of the classes Figure, ElbowHandle, ShortestDistanceConnector imply a change in behavior in one of the other classes? We suggest that these classes should change independently. Constraining the movement of line connections and finding shortest distance connections are rather separate concerns and should be developed independently of each other. The only change we could imagine, that would benefit from a redistribution of responsibilities into figure and its subclasses, would be if the area that allows for connection would have a non-rectangular shape and x and y coordinates would have to be constrained simultaneously. Consider for example allowing a non-rectangular area in an ellipsis but an ellipsis itself as shape of the connectable area. Such a change might be worth considering, but given the current state of the software variations in the strategies to manipulate line connections or to find connections are more probable, so that the responsibility is well placed in the current class.

To summarize, the lifting refactoring is obviously not an option. The pushing refactoring does need some discussion, but we are convinced enough that the responsibilities are currently already well distributed.

A3) For the same reasons as for Rectangle the coupling to Insets is not harmful, the understandability is not challenged, and encapsulation is not really broken. "Lifting forward" would increase the size of the interface Figure too much and "pushing back" would misplace a responsibility. So the 6 accesses to fields of Insets do not represent a bad smell and are thus false positives.

I1) We suggest to give the similar character of the two classes Rectangle and Insets the name "Data Class". A "Data Class" is a record-like class containing mainly state and almost no behavior.[24] I2) All accesses and calls to "Data Classes" in the "Law of Demeter" should then be ignored. I3) Besides the classes Insets and Rectangle there are a few more classes that can be considered to be "Data Classes": Color, Dimension, FontMetrics, Point and Polygon. All these classes are in the package java.awt. Enumerating these class names is an unambiguous way to define the extension of the concept "Data Class".

I4) Let us review our discussion to verify that it can be generalized from Rectangle and Insets to "Data Classes" as they are used in JHotDraw 5.1. As for the given method, one can expect that a "Data Class" is not detrimental to understandability. At least the "Data Classes" that we listed are as well stable so that the coupling toward them is not harmful. In the context of our application a "friend" class returning a "Data Class" does not necessarily reveal its internal structure, but just reports about some of its properties. The access is thus not harmful with respect to understandability, coupling and encapsulation. "Lifting" refactorings would enlarge the interface of a "friend" class without providing any benefit. The discussion of the "pushing" refactoring depended on the shape of the responsibility to push and the expected evolutionary changes in the classes under consideration. Reviewing the uses of "Data Classes" in JHotDraw 5.1, we find mostly rather trivial cases (like in `TriangleFigure.polygon()`, table 1) and the discussion above is more complex than in most other cases. This results in high enough confidence to rule out as well the "pushing" refactoring and to consider all "Data Classes" to be "everybody's friend".

---

[23] One would constrain the x and y coordinates of one rectangle to the range of the x and y coordinates of the other rectangle and finally take their arithmetic middle. This involves some extra comparisons and some extra computations that are not necessary in the original definition.

[24] See footnote 13 for a few further remarks on "Data Classes".

I5) For the given method this exception already reduces the number of violations again by 6 and we reached a reduction from 21 potential violations down to 11. The overall number of potential violations can be reduced by 539 cases or 44.4 % of the original violations.[25] The reductions per class are: `Rectangle`=366, `Point`=56, `Dimension`=36, `Color`=35, `Insets`=34, `Polygon`=7, `FontMetrics`=5.

**(D3)  Refactoring: Lift access to figure forward.**   As we are now able to put aside the obviously irrelevant accesses to the two "Data Classes", let us explore whether the remaining eleven violations in table 3 are meaningful. We will discuss why `LineConnection` and `Figure` are not recognized as "friend" classes and whether they should be seen as such below in section A.6. From the code we understand that the `ElbowHandle` is owned by a `LineConnection`, which has a `Figure` as start and as end. The geometry of these two figures is used in the calculation. If we look closer, another type is involved: `line.start()` returns an object of type `Connector`, which seems to be owned by the start figure, as the result of a call to `owner()` is stored in the local variable `startFigure`. The expression "line.end().owner()" in the next line uses as well the class `Connector`.

We discuss calls from `ElbowHandle` to `Connectors` received from an `ElbowConnection`. A1) The coupling toward `Connector` has low intensity, since only one method is called. Since the interface is from the package `CH.ifa.draw.framework`, one of the two most foundational packages of JHotDraw 5.1. it can be expected to be rather stable. So, the coupling is not very harmful. While it is not difficult to understand that a line connection connects a start and an end figure as all these objects have a visible representation, the role of the connector is not clear at all. Is the connector something visible as well? Judging by its name it could be as well a controller. We might wonder also, whether the start and end figures of line connections always have to be accessed via connectors. To access the rectangular coordinates that are used in the actual calculation we need to follow from a `LineConnection` to a `Connector` to a `Figure`, which contains the `Rectangle`. At least the intermediate step through the `Connector` class creates an additional cognitive burden[26] and seems to reveal the internal structure of `LineConnection`, thus breaking its encapsulation. The current design challenges understandability and encapsulation.

---

[25] The relative reduction in the 12 methods with the highest number of potential violations is even higher. Only 88 of 302 potential violations remain, a reduction by 70.9 %. See table 1.

A2) If we explore the class `LineConnection` we can find, that the author of this class did not mean all users of the class to follow the path along the `Connector`. We can easily relieve ourselves from this burden. `LineConnection.startFigure()` hides the detour through the `Connector` and gives us direct access to the start figure. The same holds for the end figure. "Lifting forward" was already prepared and just needs to be executed. See figure 7 for the relations of the classes. One might even consider to "push" the responsibility to constrain the coordinates to the `LineConnection`, the `Connector` or the `Figure` itself. This would allow to vary the constraining behavior depending on the actual figure or depending on the actual connecting strategy. Currently the start and end points are restricted depending on the insets, i.e. a rectangle that is completely within the figure. This might be too restrictive for ellipses and rounded rectangles. Both refactorings are considerable, but "lifting" is preferable since it is already prepared and the evolution scenarios described for "pushing" are far from the current implementation.

A3) For the call of `Connector.owner()` on the result of `line.start()` it should have become clear that 1) this is exactly the kind of dependency on internals of "non-friend" class the "Law of Demeter" discourages and that 2) this is indeed an unnecessary complexity hindering understanding and creating unnecessary obstacles to the further evolution of the relation between `LineConnection`, `Connector` and `Figure`. Lifting the access to `Connector.owner()` by using the methods `startFigure()` and `endFigure()` resolves two true violations in `constrainX(int)`, two in `constrainY(int)` and two more in `ElbowConnection.updatePoints()`.

**(D4) "Globally Accessible Member"** Let us now discuss the two calls to `range(int, int, int)` in the expressions "`Geom.range(r1x, r1x + r1width, x)`" and the other similar expression at the end of the method `constrainX(int)` in table 3. The method `range(int, int, int)` is declared `public` and `static` so that no object instance is required to call it. Consequently there is no immediate way to apply the "Law". Accesses to static members don't need references so there is no natural relation to the types of fields, parameters, instantiations.

We discuss a call from `ElbowHandle` to a static method of `Geom`. No object of class `Geom` was created. A1) The coupling is therefore rather directed immediately to the method than to the class. The provided functionality is very abstract. The class is stable since it is located in the most fundamental package of JHotDraw 5.1, namely CH.ifa.draw.util. The coupling is not harmful. The method range(int, int, int) realizes a well defined mathematical function, independent of any state, that is straightforward to understand. Since there was no reference retrieved through any "friend" class, there is no class, whose encapsulation could be broken. To sum it up, we did not find any harm. A2) Since there was no reference received through an intermediate "friend" type, the lifting refactoring can not be applied. For the same reason the pushing refactoring is not applicable. Since the method range(int, int, int) does not depend on any external state, we could consider to inline the method. This would replace the one method invocation by a sequence of two if-statements. Although this would result in one method name less, it would increase the length of the code, and the sequence would appear a few times in the code. None of the mentioned refactorings improves the code. A3) Since the potential violation is not harmful and can not be improved we consider it to be not a true positive. Besides considering it a false positive, one might even argue that that law is not even applicable to the given case.[27]

---

[26] Let us quickly discuss, whether there is a cognitive burden or not. An experienced software developer might be used to finding an additional layer of indirection in the source code. Since the indirection here is an example of a well known principle of adding additional indirections for having more flexibility, the experienced developer might not recognize any cognitive burden here. There is nevertheless a difference between a cognitive burden being unnoticed and being actually absent. Even the experienced developer needs to make a higher cognitive effort to keep four types in mind than just three. In addition the type `Connector` is not even mentioned in this method explicitly and its role is less obvious than for the three classes having an obvious geometrical meaning. The author of this work *knew* about the cognitive burden immediately after inspecting the potential violation, but he only *felt* it after removing the burden by creating a refactored version of the method.

I1) Besides range(int, int, int) being a nice method, our discussion was dominated by the fact that it is a "Globally Accessible Member". I2) Since the global accessibility is independent of the type that contains the member as well of the type of the member, we suggest to adapt the "Law" on a finer granularity and consider "Globally Accessible Members" to be "everybody's friend" without taking types into account. I3) All members that are public and static fall under this idea I4) and since they are not accessed through an intermediate "friend" type, the discussion above applies to all of them. I5) The number of potential violations is reduced by 270 cases[28] or 22.2 % of the original violations.[29]

---

[27] In the C++ version of the "Law of Demeter" [22] accesses to members of "classes of global variables used in [the method] M" were allowed as well. In Java there is no concept of a global variable. But, since public static fields can be accessed from everywhere, they come closest. The C++ rule does not explicitly allow the access to the global variables but only to members of this type. Here we do need the access to public static fields and methods themselves but not the access to members of public static fields.

[28] The reductions per class are: java.lang.Math = 67 (calls to max(int,int) = 21, min(int,int) = 9, access to PI = 8, and others), CH.ifa.draw.util.Geom = 47, CH.ifa.draw.standard.RelativeLocator = 38, java.awt.Color = 32 (as well ruled out as usage of a "Data Class"), java.lang.System = 14, CH.ifa.draw.util.ColorMap = 11, and less than 9 usages per class to members of the following classes: from JHotDraw AttributeFigure, BoxHandleKit, Clipboard, FigureChangeEventMulticaster, FigureSelection, Handle, Iconkit, LocatorConnector, PolyLineFigure, PolygonFigure; from java.awt: Cursor, Font, KeyEvent, Toolkit; from java.lang: Class, Double, Integer, Long. We did not rule out the access to all of the members of these classes but only the accesses to globally accessible members.

[29] The class CH.ifa.draw.util.Geom contains the comment "Some geometric utilities.". A first idea could therefore be to take the concept of "Utility Classes" into account, i.e., classes that are never instantiated but contain some static methods as utilities. But it is true for every publicly accessible member (fields and methods!) that we would never instantiate a class just to access their static members. So, we need to exclude all "Globally Accessible

■ **Listing 4**   Extract from the class StandardDrawingView in JHotDraw 5.1 (15 of 685 lines)

```
1   public class StandardDrawingView extends Panel /* [...] */
2   {
3       transient private Vector fSelectionHandles;
4
5       private Enumeration selectionHandles() {
6           // [ ... initialize handles if required ... ]
7           return fSelectionHandles.elements();
8       }
9
10      public void drawHandles(Graphics g) {
11          Enumeration k = selectionHandles();
12          while (k.hasMoreElements())
13              ((Handle) k.nextElement()).draw(g);
14      }
15
16  }
```

The three potential violations in the method drawHandles(Graphics) should be considered to be false positives. As discussed in (D5), working with the methods of the interface java.util.Enumeration should be considered to be an essential part of programming in Java. The call to the method draw(Graphics) of the class Handle is discussed in (D6). The field fSelectionHandles realizes a one to many association from StandardDrawingView to Handle. The field thus makes Vector as well as Handle a "friend" type.

### A.3  Observations in StandardDrawingView.drawHandles(Graphics)

The method drawHandles(Graphics) in the class StandardDrawingView as presented in table 4 draws the handles for the figures that are selected in the view. An enumeration of these handles is provided by the method selectionHandles() in the same class. The code contains three potential violations. (D5) discusses the calls to hasMoreElements() and nextElement() on the Enumeration k and (D6) discusses the call to draw(Graphics) on the elements of the enumeration.

**(D5) "Collection Types"**   We start the discussion of the code in table 4 with the calls to the methods hasMoreElements() and nextElement() on the Enumeration k received from the call to selectionHandles(). The two methods already amount to the full interface java.util.Enumeration.

We discuss calls from StandardDrawingView to an Enumeration received from the calling class itself. A1) The coupling toward Enumeration has low intensity since only two methods are called. These are all methods of the interface. Enumeration is — as an external type of the JDK — very stable and the coupling is not harmful. The interface is an essential part of Java. Every programmer needs to learn how to work with enumerations and since Java 1.2 (JHotDraw 5.1 was developed with Java 1.1) with

---

Members" from the "Law" anyway. Whether and where a software system contains "Utility Classes" is nevertheless an interesting information in its own right.

collections in general. Collections—like arrays—can hold many objects but have more freedom in how the objects are stored and accessed. A field of type `Enumeration` realizes what is modeled in UML by the multiplicity of an association. In the given example the use of `Enumeration` does not break encapsulation for two reasons. First, it does not reveal anything about `StandardDrawingView`, on the contrary, the `Enumeration` gives a restricted access to the elements in the `Vector` of handles. Second, we happen to access the enumeration from the `StandardDrawingView` itself. Working with `Enumerations` is an essential part of programming in Java and the access to its methods therefore unproblematic. A2) Lifting the access would require to have two instead of one access method to the selection handles, since we would need as well an accessor of the individual handles. Pushing had the same negative effect since it would mean to iterate directly through the original vector of selected handles in the same class. Neither the "lifting" refactoring nor the "pushing" refactoring is helpful here. A3) Using enumerations had been a good practice in Java 1.1 and using collections became standard—since Java 1.2—and type-safe since Java 1.5. It must be possible to access these types. These potential violations are false positive.

I1) "Collection Types" are essentially part of the Java language. They offer the standard way to store multiple objects together. They are preferred over arrays, when time and memory performance is not the main concern. I2) Hence, accessing the members of "Collection Types" should always be possible. These types should be considered "everybody's friend". I3) All types that implement one of the interfaces `Enumeration`, `Map` or `Set` from the package java.util or the interface java.lang.Iterable should be considered "Collection Types" and "friends" of all classes.[30] I4) The motivation for the adaptation is independent of the given case and can thus be generalized. The definition of the extension is flexible and covers e.g. as well `FigureEnumeration`. It is vulnerable against abuses of the interfaces for classes realizing other responsibilities than just collecting objects. I5) With the suggested adaptation 137 potential violations can be identified as false positives and removed.

**(D6) "Known Type in Aggregation"**    The last line of `drawHandles(Graphics)` in the class `StandardDrawingView` presented in table 4 takes an object from the enumeration and invokes the method `draw(Graphics)` on it, which requires casting the object to `Handle`.[31] `StandardDrawingView` has a one to many association to the class `Handle`, which is realized through the field `fSelectionHandles` of type `Vector`.

We discuss calls from `StandardDrawingView` to `Handles` received from an `Enumeration` which contains the elements of a field of type `Vector`. Fields of type array, `Vector` and—since Java 1.2—of collections are Java's way to express associations with multiplicity higher than 1. We argued in (D5) for collections that they are just a means for realizing these associations. The even more essential aspect of a field are the objects contained in it and their type. Unfortunately up to Java 1.5 there was no way to express, what type is contained in a collection. This is what makes the cast necessary. Otherwise we would not be looking at a potential violation here. The fact that the type of the objects contained in the `Vector` `fSelectionHandles` can not be expressed directly is really unfortunate. To verify that the cast to `Handle` is safe, i.e. that all objects in the respective `Enumeration` are indeed always at least of type `Handle` one would need to review all possible flows of objects into the `Enumeration` and before into the `Vector`.

---

[30] With respect to JHotDraw 5.1. the interface `Enumeration` and the class `Vector` would have been enough, since the Collection Framework was only introduced in Java 1.2. The core of the idea had nevertheless already been present.

[31] A cast in Java never changes any object. The cast just assures the compiler of the type the object has. When this assurance turns out to be not fulfilled at runtime, a `ClassCastException` is thrown.

A1) How do we evaluate the call to `draw(Graphics)` with respect to understandability, coupling and encapsulation, given that the field `fSelectionHandles` realizes the one to many association to `Handle`? Understanding that the field contains `Handles` and only `Handles` is not immediate, since this information is not located at one place. Once we know the type in the aggregation, understandability is no issue anymore, the coupling from `StandardDrawingView` to `Handle` is a natural consequence and encapsulation of `StandardDrawingView` is not broken by accessing `Handle`. The current design is not at all bad. What is missing are the Generic Types introduced in Java 1.5.

A2) Both, `Vector` and `Enumeration`, are meant to be very general types and should not contain a method that is specific to JHotDraw in their interface (after a "lifting" refactoring) or accept a specific responsibility (after a "pushing" refactoring). If we wanted to apply one of the two refactorings under consideration, we would need to create a new specific type that contains a collection of `Handles`. Let us call this potential type `HandleEnumeration`. Since this type would collect more than one `Handle`, we could not lift the draw method of a specific handle to this type. Conversely it would be very easy to push the responsibility to draw all handles to this potential type into a method `drawAll(Graphics)`. Given that a push refactoring would be possible, we need to discuss, whether the addition of the type `HandleEnumeration` makes sense.[32] The main question is, whether the type pulls it weight. To answer this question we reviewed `StandardDrawingView` looking for other responsibilities to push. This type could as well accept the responsibility of the method `findHandle(int, int)` of the `StandardDrawingView`. Besides this, only two methods for emptying the enumeration and for adding one or more handles would be required. The type would just be used inside `StandardDrawingView`, its interface would just be designed to be used by the view. We suggest not to introduce such a type. The responsibilities are not clear enough. The methods for maintaining its state would just forward to a `Vector` inside. In addition, one would need to create a similar type as well for the selected Figures[33] managed in the field fSelection. Managing these two collections and maintaining its consistency is the well defined main responsibility of the StandardDrawingView and splitting this responsibility into parts does not clarify the design, but separates aspects that belong together.

---

[32] Lieberherr and Holland explore for a given example [21, Figure 7] the option to have a specific class to realize a list and to push responsibilities into this class. In the context of that paper the example has the function to illustrate that a "lifting" refactoring is not always possible so that the "pushing" refactoring is sometimes required. From the subjective observation of the author of this work, developers do not tend to implement collections with functionality that is specific to the elements in the collection anymore. As the mentioned example shows, this has been different before. Technology favoring collections with no element specific functionality would be the Standard Template Library for C++ or the Collections Framework for Java.

A3) Once we accept the design decision to maintain the one to many associations of the StandardDrawingView to the selected figures and their handles using the rather general type Vector and Enumeration, there is no option to "lift" or to "push". The discussion of understandability, coupling and encapsulation led as well to the conclusion to consider the given potential violation to be a false positive.

I1) We call this idea "Known Type in Aggregation". In Java 1.5 we could express the known type to the compiler by using the type Vector<Handle>, but till then we need to I2) adapt the "Law of Demeter" to make the known type available to the analysis. I3) The idea refers to all fields of type Vector. In almost all cases, we can infer the types aggregated immediately from the types of objects added to the field via the method addElement(Object). Only for the given case of the field fSelectionHandles the inference is not that immediate,[34] so that we either would need to add a slightly more involved type-flow analysis or add the knowledge for the singular case manually. I4) This heuristic is good enough for JHotDraw 5.1. For other software one might need to refer to further methods. The decision to consider all types of objects stored in fields of type Vector to be "friends" is clearly in the spirit of the "Law of Demeter". I5) This adaptation removes 40 false positives.

### A.4 Observations in IconKit.loadImageResource(String)

The method loadImageResource(String) in the class IconKit as presented in table 5 retrieves the URL of a resource based on the location of the class file. With the help of the only instance of the class Toolkit it creates an Image from this URL.

After taking the discussions in the previous sections into account, the method contains four potential violations. The call to getResource(String) on the result of getClass() will be discussed in (D7). The statement "System.out.println(resourcename)" will be discussed in (D8). The call to getContent() on the URL url will be discussed in (D9). The call to createImage(ImageProducer) on the Toolkit toolkit retrieved in the first line of the method through the "Singleton Access" "Toolkit.getDefaultToolkit()" will be discussed in (D10). Finally similar cases to the latter will be discussed in (D11) without explicit code.

---

[33] The already existing type FigureSelection does not provide the required functionality. It is designed to pass a selection of figures through the clipboard. This type is indeed used by a few other classes.

[34] The field is only filled in the method selectionHandles(). The inference would be as follows: Figure.handles() returns a Vector of Handles. The Vector.elements() returns an Enumeration providing access to the same Handles via Enumeration.nextElement(), which are stored in fSelectionHandles via Vector.addElement(Object).

■ **Listing 5** Extract from the class Iconkit in JHotDraw 5.1 (17 of 140 lines)

```
1   public class Iconkit {
2
3       private static boolean fgDebug = false;
4
5       public Image loadImageResource(String resourcename) {
6           Toolkit toolkit = Toolkit.getDefaultToolkit();
7           try {
8               java.net.URL url = getClass().getResource(resourcename);
9               if (fgDebug)
10                  System.out.println(resourcename);
11              return toolkit.createImage((ImageProducer) url.getContent());
12          } catch (Exception ex) {
13              return null;
14          }
15      }
16
17  }
```

We had already suggested in (D4) to tolerate calls to "Globally Accessible Members" like Sytem.out and getDefaultToolkit(). Since the latter call is an access to a "Singleton" it plays the same role as an instantiation of the class Toolkit. As discussed in (D10) Toolkit is thus as well a "friend" type and the call to createImage(ImageProducer) is consequently a false positive. The discussion in (D7) leads to the suggestion to consider all classes in the package java.lang to be "everybody's friend", so that the call to Class.getResource(String) is as well a false positive. The same would hold for potential usages of the exception ex. (D8) suggests to tolerate calls to print methods on System.out and (D9) suggests to consider java.net.URL to be a "Data Class". So, none of the shown potential violations prevails.

**(D7) "Types of the Java Language"**   We discuss the call to the method getResource(String) of the class Class. Motivated by the previous discussion, we will focus on the aspect that the class is in the package java.lang and thus part of the language.

We discuss calls from Iconkit to an object of class Class received from the superclass Object. A1) The coupling toward Class has low intensity, since only one method is called. Since java.lang.Class is in addition as an external class of the JDK very stable, the coupling is not harmful. The intention of the method to load a resource relatively located to a class is clear.[35] Since no internals of Object are revealed, encapsulation is not broken. The call is thus not problematic. A2) In the given situation no refactoring is possible, since Object is an external class. Let us nevertheless assume that we could change this class: A lifted method Object.getResource(String) would be too rarely used to be placed in the most general classes of all. The functionality is unrelated to an instance but to its class. Thus it is helpful to have this method in the class Class. The maximal responsibility that we could push would be again just the call to getResource(String) so that pushing would be the same as as lifting. No refactoring is possible and not even recommendable. A3) We consider the given potential violation to be a false positive.

---

[35] Identifying the actual location of the resource and ensuring that it gets deployed is sometimes difficult, but that is unrelated to the understandability of the code.

I1) In the paragraph before we had argued that "Collection Types" should be considered as part of the language. The same arguments holds a fortiori for classes in the package java.lang and we suggest to summarize all of them under the name "Types of the Java Language" and I2) consider all "Types of the Java Language" to be "everybody's friend". I3) These types can be identified by their package java.lang. I4) Since these classes are so fundamental for Java, our confidence is high enough. I5) This adaptation removes 128 potential violations[36] of which 94 had already been covered by the adaptation given in (D4) "Globally Accessible Member".

**(D8) "Standard Printing Idiom"** Next we discuss the call to the method println(String) in the class java.io.PrintStream in the statement System.out.println(resourcename).

We discuss calls from Iconkit to a PrintStream accessed via the static field System.out. A1) The coupling toward PrintStream has low intensity, since only one method is called. Since java.io.PrintStream is in addition as an external class of the JDK very stable, the coupling is not harmful. Since the given statement is the central line of most "Hello World!"-programs in Java, i.e. the first program a developer ever writes to see that a program in a certain language can actually communicate with the outside, it is easy understandable. Since the PrintStream is accessed through the "Globally Accessible Member" System.out there is no intermediate "friend", whose encapsulation could have been broken. Overall the statement is unproblematic.

A2) Since there is no intermediate "friend" we have no class into which to lift or to push. One might consider the introduction of a new class like a Logger. But, this is only reasonable if the goal is to fulfill further intentions like getting more control over the printing process or deciding what to print when and where. We discussed the option to create a "friend" class already in (D5). Here again, an extra class does not pull its weight. Just from the perspective of the "Law of Demeter" no refactoring makes sense.

A3) Since the coupling is toward a stable class, the code trivial, no encapsulation broken and no refactoring available, the given potential violation is a false positive.

I1) We are looking at the most basic approach to create output in Java - at the "Standard Printing Idiom". I2) The most narrow adaptation would be to allow access to members (!) used for standard printing, if System.out was accessed. I3) These methods can be identfied by name pattern and type: "PrintStream.print*()". I4) Since our argument did not use specific context of the potential violation our confidence in the adaptation is high. I5) This resolves eleven similar cases.

**(D9) java.net.URL as "Data Class"** In table 5 the object url of type java.net.URL was retrieved through a call to getResource(String). Since there is no other relation to URL the call to getContent() is a potential violation.

We discuss calls from Iconkit to a URL received from a Class. A1) The coupling toward URL has low intensity, since only one method is called. Since java.net.URL is in addition as an external class of the JDK very stable, the coupling is not harmful. Since a URL is a common concept, it is easy to understand. The call to getContent() does not reveal internals of the class but provides a service based on the value of the URL. Overall, there is no problem with the call. A2) The "pushing" and "lifting" refactorings are impossible since the class Class is external. A method with the signature Toolkit.createImage(URL) would be useful and is indeed available in later versions of Java, but in the JDK 1.1. it is missing.[37] The current design can not be improved by refactorings. A3) We suggest to consider the given potential violation to be a false positive since we are looking at a stable class providing a common concept, where no encapsulation is violated and no refactoring is possible.

---

[36] The accessed classes are the value types Boolean, Double, Integer, Long and String, the fundamental classes Class, System and Throwable and the class Math.

I1) Since a URL is rather a value than an active object, we suggest to subsume the class URL under the design idea "Data Class" as introduced in (D2). I2) We had already suggested to consider "Data Classes" to be "everybody's friend" I3) We change the extension of the design idea by adding the class java.net.URL to it. I4) The class URL seems to be a bit more active than the "Data Classes" that we have seen before, but our confidence is still high enough to extend the idea and the adaptation to the given case, especially since I5) we are discussing just this single case and there is no other similar case in JHotDraw 5.1.

**(D10) "Singleton" Design Pattern**    Finally, the extract from Iconkit in Figure 5 contains the interesting case of the call createImage(ImageProducer) on toolkit. The "Law of Demeter" explicitly allows calls to classes that are instantiated in a method, but Toolkit is a case of the "singleton" design pattern [13, pages 127–138] and therefore the only instance of the class is accessed through a specific method. Here this method is named getDefaultToolkit().

We discuss calls from Iconkit to a Toolkit received from a static method of Toolkit. Since the "Singleton" pattern allows no other way to instantiate the Toolkit, considering the "Singleton Access" to play the same role as regular instantiation in the "Law of Demeter" strongly suggests itself. Let us nevertheless quickly examine the case.

A1) The coupling toward Toolkit has low intensity, since only one method is called. Since java.awt.Toolkit is in addition an external class of the JDK very stable, the coupling is not harmful. The call itself has the simple purpose of creating an image from an appropriate representation. What confuses the method a bit are the crosscutting concerns of logging and exception swallowing entangled in the method. The instance of class Toolkit is not received through an intermediate class whose encapsulation could have been broken. The call to createImage(ImageProducer) is thus unproblematic. A2) Since there is no intermediate class, neither the lifting nor the pushing refactoring is possible. One could consider creating a static method in Toolkit with the same functionality as loadImageResource(String) but this would contradict the intention of having a (single) instance of the class and the overall distribution of responsibilities between these classes. Removing the potential violation by refactoring will not lead to consistent code. A3) The current code is overall in good shape. No refactoring would resolve the potential violation. Most importantly, the "Singleton Access" it conceptually the same as an instantiation, so that we consider the given potential violation a false positive.

---

[37] See e.g. https://www.cs.princeton.edu/courses/archive/fall97/cs461/jdkdocs/api/java.awt.Toolkit.html

I1) The idea of a "Singleton Access" is thus central for our adaptation. I2) The "Singleton Access" "introduces" the accessed type to the method, so that it becomes a "friend" type. I3) Although there are good chances to detect Singletons using a heuristic, we suggest to refer to the classes by name to remove any ambiguity. The accessor can then be inferred. In JHotDraw 5.1. we find the following "Singletons": Toolkit with its accessor getDefaultToolkit(), IconKit with its accessor instance(), Clipboard with its accessor getClipboard(). I4) Based on our discussion our confidence in the suggested adaptation is very high and the confidence in the definition of the extension is as well high enough. I5) Overall this adaptation removed 24 potential violations.

**(D11) Further Creational Design Patterns** There are other method calls that are conceptually the same as instantiations. In JHotDraw 5.1 we find two variations of a "Factory Method". The method print() of the class DrawApplication contains the expression "getToolkit().getPrintJob(this, "Print Drawing", null)", where getToolkit() is an inherited method and getPrintJob() is an abstract method with an implementation in a subclass of Toolkit. There is just this one call on the Toolkit object and two calls on the created PrintJob object.

When we discuss the call from DrawApplication to a Toolkit received from the superclass Window or when we discuss the calls from DrawApplication to a PrintJob received from that Toolkit, we happen to find again no reason for improvement: A1) The coupling has low intensity (one method called on Toolkit and two methods called on PrintJob) and both types are very stable (package java.awt) so that the coupling is not harmful. The methods are used for an easy to understand typical printing process. Just from the name of the methods it is hard to judge, whether they reveal hidden information and thus break encapsulation. As soon as we know that the methods are "Factory Methods", we understand that they do not provide a part but create new objects that are meant to be used independently. The calls are in terms of coupling, understandability and encapsulation unproblematic. A2) In the given case the "lifting" as well as the "pushing" refactoring happen to be impossible since Window as well as Toolkit are external classes. A3) Like in (D10) it is nice to know that the potential violations that we are looking at are not problematic, but that is not essential. The decisive argument is, that calls to these methods delivering instances are indeed conceptually the same as instantiation. We thus consider the potential violations to be false positives.

I1) We subsume the two methods under the design idea "Factory Method" and I2) consider the calls to these methods to be like instantiation, i.e. the return types are "friends" where the methods are called. I3) We suggest to identify the "Factory Methods" just by enumerating their names. I4) The reason for the adaptation is the conceptual equivalence of instantiation and calls to "Factory Methods". This is independent of the given case. I5) JHotDraw 5.1 contains just the discussed three calls on objects received from "Factory Methods" potentially violating the "Law".

The "Factory Method" in the sense of the "Factory Method Pattern" [13, pages 107–116] is as well a "Hook Operation" of a "Template Method" typically allowing frameworks to defer the decision which type to instantiate to concrete application classes. getToolkit() is declared in the class java.awt.Component and overridden in the class java.awt.Window and could have been overridden as well in DrawApplication. For the full pattern, we are nevertheless missing the "Template Method" in a framework class. getPrintJob() is more of a "Factory Method" in the sense of the "Abstract Factory Pattern" [13, pages 87–95], where the role of an "Abstract Factory" is superimposed onto Toolkit. Furthermore, Kerievsky [17] distinguishes "Creation Methods" from "Factory Methods". The terminology varies in the literature though. For example one could narrow the name "Creation Method" down to static methods returning instances of the same class. Being too rigid about terminology hampers however the discourse about ideas. In this work we consider a "Creation Method" to be any method with the main purpose of creating objects.

### A.5 Observations in JavaDrawApp.createWindowMenu()

The method createWindowMenu() in the class JavaDrawApp as presented in table 6 creates one of the menus of the application. It adds to this menu one item that offers the user the possibility to open a new window. The ability to open a window is implemented in the method openView() of the class JavaDrawApp. The menu item is connected to this method by means of an anonymous inner class implementing the ActionListener interface. The method contains only one potential violation, namely the call to openView() which is discussed in (D12).

**(D12) Inner classes share the friends of the outer class**   The extract from the class JavaDrawApp in table 6 shows a type of violation that occurs only 23 times in the project, but might give a developer the impression that the "Law of Demeter" is not well suited for Java programs. The call to openView() from the method actionPerformed(ActionEvent) is a potential violation, because it is a call from one class to another class that is not related to the first class in one of the ways listed in the definition of the "Law". The anonymous class implementing the interface ActionListener is an inner class of the outer class JavaDrawApp. An instance of this anonymous inner class belongs to an instance of the outer class and has access to it.

We discuss calls from an anonymous inner class to its outer class JavaDrawApp which is by language design accessible from the inner class. A1) The coupling from the inner class to the outer class is low here. The maintenance effort created by a change to the outer class requiring changes in the inner class is anyway moderate since both are located close to each other. Since the anonymous inner class and the called method of the outer class contribute to the same or at least closely related responsibilities they can be expected to be of similar stability. The construction "anonymous inner class" is one of the more advanced Java concepts and requires thus some initial learning effort. The syntax for calling methods of the outer class, accessing its fields or the parameters for the enclosing method is however very simple. All these members can be accessed as if they were own members. This means also that the structure of the outer class

**Did JHotDraw Respect the Law of Good Style?**

■ **Listing 6**  Extract from the class `JavaDrawApp` in JHotDraw 5.1 (20 of 180 lines)

```
1   public class JavaDrawApp extends DrawApplication {
2
3       protected Menu createWindowMenu() {
4           Menu menu = new Menu("Window");
5           MenuItem mi = new MenuItem("New Window");
6           mi.addActionListener(
7               new ActionListener() {
8                   public void actionPerformed(ActionEvent event) {
9                       openView();
10                  }
11              }
12          );
13          menu.add(mi);
14          return menu;
15      }
16
17      public void openView() {
18          // [ ... open a new window ... ]
19      }
20
21  }
```

The instantiation "`new ActionListener(){ [...] }`" creates implicitly a new class. This class has no name. Its instance has access to the instance of the enclosing class `JavaDrawApp` on which the method `createWindowMenu()` was called. The creation of such anonymous inner classes is in Java the idiomatic way to add a listener. Since the method `openView()` is not in the anonymous inner class, the call is a potential violation. As discussed in (D12) accesses to the outer class are the very reason, why the construct of anonymous inner classes exists. We therefore suggest to consider every type that is a "friend" for the enclosing method to be a "friend" of the inner class.

is anyway not hidden from the inner class so that no encapsulation is affected. The (implicit) reference is also not received through a third object, whose encapsulation could have been broken. The construct itself is not that easy, but an essential part of programming in Java. Given the construct, accesses to the outer class, its fields and the parameters of the method are unproblematic. A2) Pushing and lifting are in the given example synonymous since there is only the one call. They are meaningless, since there is no intermediate class. If the inner class would access a field of the outer class, lifting might become possible. Pushing might become possible, if there is more code in the anonymous inner class. Having too much code in an anonymous inner class is not an question of, which objects to access but simply of size. In JHotDraw 5.1 anonymous inner classes are short. There is no indication for a refactoring. A3) The given potential violation is thus a false positive.

I1) The relevant concept here is the concept of an "Anonymous Inner Class". I2) We suggest to consider all "friends" of the enclosing method to be as well "friends" of the "Anonymous Inner Classes". I3) "Anonymous Inner Classes" can be identified as parts of an instantiation expression. I4) Since the access from inner to outer classes is essential for this language feature, we are confident that the decision for the given case can

■ **Listing 7**  Extract from the class RadiusHandle in JHotDraw 5.1 (17 of 57 lines)

```
1  class RadiusHandle extends AbstractHandle {
2
3      private RoundRectangleFigure fOwner;
4      private static final int OFFSET = 4;
5
6      public RadiusHandle(RoundRectangleFigure owner) {
7          super(owner);
8          fOwner = owner;
9      }
10
11     public Point locate() {
12         Point radius = fOwner.getArc();
13         Rectangle r = fOwner.displayBox();
14         return new Point(r.x+radius.x/2+OFFSET, r.y+radius.y/2+OFFSET);
15     }
16
17 }
```

RadiusHandle maintains a field fOwner of the specific type RoundRectangleFigure. The type of the constructor parameter ensures the specific type. See figure 8 for a partial class diagram. The method locate() can use the method getArc() declared in RoundRectangleFigure as safely as the method displayBox() declared in Figure. In contrast to TriangleRotationHandle presented in table 8 no cast is necessary. Accesses to Point or Rectangle fields like in the return statement are considered false positives since (D2).

be generalized to other "Anonymous Inner Classes". I5) This adaptation removes 23 potential violations, that would be otherwise very annoying to any Java developer.

### A.6  Observations in subclasses of AbstractHandle

In section A.2 we suggested that LineConnection should be considered a "friend" type in the method constrainX(int) of the class ElbowHandle. We postponed the reasoning to the current section, because we wanted to conduct this discussion with more context. It is helpful to explore what problem is solved by the method ownerConnection() in the class ElbowHandle as presented in table 3 and how related classes solve the same problem.

Handles are meant to manipulate figures. All handles are owned by a figure and have a reference to their owning figure. The handles are displayed as small squares or circles and enable the user to manipulate the figures with the mouse.[38] The kind of manipulation provided by a handle might depend on the type of the figure. AbstractHandle realizes the association from the handle to the figure through a field of type Figure and gives access to this field via a method owner(). In figure 7 we had presented a partial class diagram showing an overview of this association for the class ElbowHandle. The ElbowHandle needs to know that its owner is not only a Figure but more specifically a `LineConnection`. This knowledge is localized in the method `ownerConnection()`. Other handle implementations represent this knowledge in other ways as will be discussed in the following.[39] For one paragraph we leave the "Law of Demeter" aside and just describe the different realizations of this association in the paragraph "Disparity of Covariant Owner Representation". Surprisingly we find five different variations. In the subsequent paragraph we will then discuss how these variations would be evaluated by the unadapted "Law of Demeter" in the paragraph "The "Law of Demeter" on the Covariant Owner Representations".

---

[38] As these handles translate mouse events (forming the interface expected by the HandleTracker tool) into the available change operations of the figures, they should be considered as "Adapters" in the sense of the "Adapter" pattern [13, pages 139–150].

Based on these reflections we will justify our decision for the access to `LineConnection` in `ElbowHandle.constrainX(int)` in (D15). Finally, there will be as well refactoring suggestions for example for `FontSizeHandle` accessing `TextFigure` (D16). Both discussions will impact as well classes that do not belong to the hierarchy of handle classes. Furthermore the considerations within the current section will turn out to be prototypical for further discussions presented later.[40]

**(D13) Disparity of Covariant Owner Representation**   To illustrate the relation between handles and figures table 7 presents an extract of the code for the `RadiusHandle` and figure 8 shows a reduced class diagram of the related classes. Figure 5 shows minimal versions of the class diagrams of `ElbowHandle` and `RadiusHandle` for direct comparison.

The `RadiusHandle` enables the user to adjust the radius of the rounded corners of `RoundRectangleFigures`. Given the close collaboration between the handles and the figures, one might argue that figures should be considered "friends" in the sense of the "Law of Demeter". Indeed, no violation is found in the case of the `RadiusHandle`. In contrast to the eight classes showing violations `RadiusHandle` has not only the inherited association to the `Figure` interface but as well a second association to the class `RoundRectangleFigure`. Since both fields are private they can have the same name, as private fields are not visible to other classes including subclasses. The two `fOwner` fields belong to the same object, but how they are used can only be described in the respective class. Both fields refer to the same object, but the field in `RadiusHandle` has the specific type so that the `RoundRectangleFigure` becomes a known "friend". In the other classes the owner reference is just downcast where needed or the downcast is localized in a specific accessor method like `ownerConnection()` as we have seen in table 3.

Table 2 elaborates the disparity of the different approaches to access the specific figure from a specific handle. It distinguishes not only the three degrees of localization of the type informations but as well, whether the constructor did declare the required specific type. The degrees of localization of the type information are: 1) The result of

---

■ **Table 2** Five different strategies to make the covariant owner type accessible. The subclasses of `AbstractHandle` have subclasses of `Figure` as owner and depend on their specific features. Only in the case of `ChangeConnectionHandle` and `RadiusHandle` the specific type information about the owner is readily available to the "Law of Demeter".

| | a) "Silent Type Expectation": The constructor parameter owner has just the type `Figure` although a more specific type is required. | b) "Explicit Type Declaration": The constructor parameter owner has the specific required subtype of `Figure`. |
|---|---|---|
| 1) "Scattered Downcasts": The owner is cast to the specific type, where required. | In `FontSizeHandle` the owner is cast to `TextFigure` twice. There is *no hint* in the class, why these casts are safe! | `PolygonScaleHandle` and `TriangleRotationHandle` take in the constructor as owner a `PolygonFigure` or `TriangleFigure` resp. The owner is then cast down to this type four or two times resp. See table 8 for an excerpt of the latter class. |
| 2) "Covariant Accessor Method": The owner is cast in an accessor method having the specific return type. | *No example.* All three classes having a "Covariant Accessor Method" have an "Explicit Type Declaration" for the constructor parameter owner. | In `ElbowHandle` the owner parameter is declared to be of type `LineConnection` and the downcast is localized in one method as can be seen in table 3. The same is true for `PolygonHandle` and its owning `PolygonFigure` as well as for `PolyLineHandle` and its owning `PolyLineFigure`. |
| 3) "Covariant Redundant Field": The owner is stored in addition in a second field having the specific type. | `ChangeConnectionHandle` casts the owner in its constructor to the required specific type `ConnectionFigure` and stores it in a field. There is *no hint* in the constructor, why this cast is safe! `ChangeConnectionStartHandle` and `ChangeConnectionEndHandle` inherit the field. | `RadiusHandle` declares its owner parameter in the constructor to be of type `RoundRectangleFigure` and stores also a reference in a field of the same type, so that no cast is required. See table 7 and figure 8. |

■ **Listing 8**   Extract from the class TriangleRotationHandle in JHotDraw 5.1 (31 of 74 lines)

```
1  class TriangleRotationHandle extends AbstractHandle {
2
3     private Point fOrigin = null;
4     private Point fCenter = null;
5
6     public TriangleRotationHandle(TriangleFigure owner) {
7        super(owner);
8     }
9
10    public void invokeStart(int x, int y, Drawing drawing) {
11       fCenter = owner().center();
12       fOrigin = getOrigin();
13    }
14
15    public void invokeStep (int dx, int dy, Drawing drawing) {
16       double angle = Math.atan2(fOrigin.y + dy - fCenter.y,
17                                 fOrigin.x + dx - fCenter.x);
18       ((TriangleFigure)(owner())).rotate(angle);
19    }
20
21    public void invokeEnd (int dx, int dy, Drawing drawing) {
22       fOrigin = null;
23       fCenter = null;
24    }
25
26    Point getOrigin() {
27       Polygon p = ((TriangleFigure)(owner())).polygon();
28       // [...]
29    }
30
31 }
```

Compare how this class handles the reference to its owning TriangleFigure to how RadiusHandle handles its reference to RoundRectangleFigure in table 7. If TriangleRotationFigure would keep as well a specialized reference to a TriangleFigure the calls to TriangleFigure.rotate(double) in the method invokeStep(int, int, Drawing) and to TriangleFigure.polygon() in the method getOrigin() would not be potential violations. The same is a fortiori true for the call to Figure.center() in the method invokeStart(int, int, Drawing). The discussion in (D15) suggests to interpret types of "Constituting Constructor Parameters" like field types, so that these three calls are as well false positives. For an overview of the different approaches to handle the owner references see table 2. The call to the "Globally Accessible Member" Geom.atan2(double, double) had already been ruled out in (D4). Notice that the accesses to the fields of Point are no potential violations, as TriangleRotationFigure has fields of type Point.

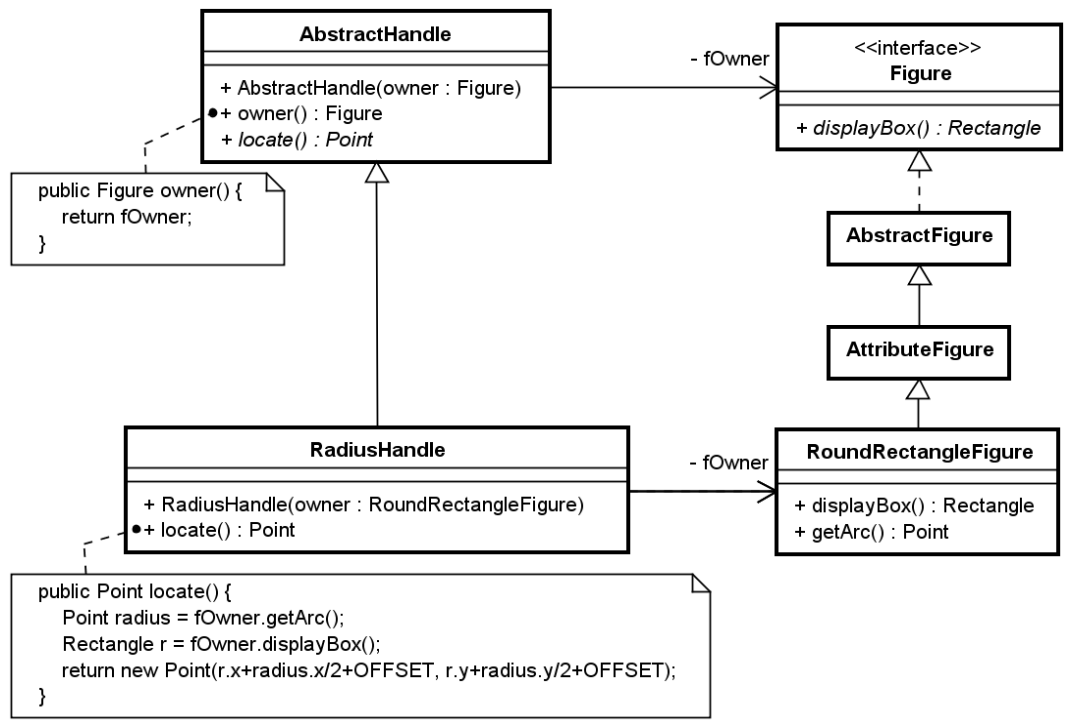

■ **Figure 8**  Partial class diagram showing the two associations from `RadiusHandle` to the `RoundRectangleFigure`. See as well the source code in table 7 and compare this class diagram to figure 7. figure 5 allows for side-by-side comparison. Ten handle classes use features from specific figure classes and not only from the Figure interface. The class `RadiusHandle` does *not* have a potential violation as it has an extra reference to its owner with the specific type, but eight classes do show potential violations. See table 2 for details.

a call to `owner()` is cast to the specific type, where required ("Scattered Downcasts"). 2) The cast on the result of `owner()` is localized in an accessor method having the specific return type ("Covariant Accessor Method"). 3) A reference to the owner is stored as well in the specific handle in a field having the required specific subtype of `Figure`. ("Covariant Redundant Field"). With respect to the constructor parameter `owner` the possibilities are: a) The constructor parameter has just the type `Figure` although a more specific type is required ("Silent Type Expectation"). b) The constructor parameter has the required specific subtype of `Figure` ("Explicit Type Declaration").

All but one of the possible combinations are indeed found in the code. The original developer might have been sure of the fact that the owner of the handle is indeed of the required type, but a new developer needs to reason about the code in five different ways to understand the type safety. In cases of a) "Silent Type Expectations" he needs to inspect as well the creators of the handles[41] to understand that the owner is indeed

_______________
[41] The flow analysis for `ChangeConnectionHandle` goes like follows: The constructor is only called from the constructors of the two subclasses `ChangeConnectionStartHandle` and `ChangeConnectionEndHandle`, which are invoked from `ElbowConnection` and

of the required specific type. In the cases of b) "Explicit Type Declaration" he can at least be sure that the handle is always instantiated with an owner of the required specific type. In the cases of 1) "Scattered Downcasts" and 2) "Covariant Accessor Methods" he needs to verify that the inherited method owner() actually returns the owner that is passed as argument to the constructor. In the cases of 3) "Covariant Redundant Fields" the developer might want to assure himself that the owner never changes, so that in fact the redundant fields are always pointing to the same object. All cases require at least a bit of reasoning and the disparity of the approaches makes it impossible for the developer to just recall one scheme of reasoning.

While the disparity is itself already a burden to the developer, some of the solutions are better than the others. Since the specific types are anyway required at runtime, there is no reason to keep the "Silent Type Expectations". A new developer might even instantiate a handle with a type that does not fulfill the expectation and hopefully experience the runtime error before a user experiences it. In general the analysis that could guarantee a specific type might get quite involved.[42] For JHotDraw 5.1 these guarantees can be verified with a bit of code inspection.[43] In cases like these, where the type can indeed be known, "Explicit Type Declarations" are thus clearly preferable.

The solution "Scattered Downcasts" is inferior to the others since one needs to reason not only for one downcast but for a few. If a type needs to change, not only one location requires change, but a few. Both difficulties might as well encourage developers not to think thoroughly about the types but to develop an attitude of downcasting, just where it is required, replacing insight with speculation. The remaining two of the six possible combinations are both reasonably good solutions. Given the "Explicit Type Declaration" in the constructor, the "Covariant Accessor Method" localizes the downcast in one place and the "Covariant Redundant Field" does not even require a cast at the cost of having redundant fields. Since in our case the owner never changes, both solutions are good. If the owner could change, the effort for keeping the redundant fields in a consistent state might be an obstacle for the latter approach.

Let us summarize the conclusions of this discussion: The solutions for having a specific owner for specific handles in JHotDraw 5.1 show unnecessary variation. "Silent Type Expectations" and "Scattered Downcasts" should be avoided. A combination of "Explicit Type Declaration" in the constructor with a "Covariant Accessor Method" or a "Covariant Redundant Field" are equally good solutions. In the next discussion we switch back to the "Law of Demeter" and review how the findings of the current discussion are related to potential violations of the "Law".

---

LineConnection — both indirectly implementing the ConnectionFigure interface. They pass a reference to themselves to the constructor, so that the owner parameter is indeed of type ConnectionFigure. Changing the type of the parameter owner to ConnectionFigure in the three constructors would allow to remove the downcast in ChangeConnectionHandle.

[42] See for example the extended discussion by Tip, Kiezun and Bäumer [42].

[43] For the reasoning for ChangeConnectionStartHandle and ChangeConnectionEndHandle see footnote 41. For FontSizeHandle one just has to consult TextFigure.handles() to see that TextFigure's this is passed as argument for owner.

**(D14) The "Law of Demeter" on the Covariant Owner Representations**    To illustrate the relation between handles and figures table 7 presented an extract of the code for the `RadiusHandle` and figure 8 showed a reduced class diagram of the related classes. In contrast to `ElbowHandle` (see table 3 and figure 7) and `TriangleRotationHandle` (see table 8) this class has no potential violations with respect to the access to the owner. In total eight handle classes contain a potential violation where they access their owning figure.

Our initial impression that the type of the owner should be considered to be a "friend" type is supported by the observation, that the conceptual relation between the handle object and the figure object are always the same. Only the "Covariant Redundant Field" provides the type information in a way that is recognized by the definition of the "Law of Demeter". For the other cases we need to discuss, if and how we can decide to consider the specific figures to be "friends" as well.

The given potential violations are in any case interesting, since they point us to the problematic "Silent Type Expectations" and "Scattered Downcasts". If one adapts the "Law of Demeter" to tolerate these cases, one might want to establish other complementary heuristics that capture these two problems. Let us now discuss when to adapt the "Law of Demeter" and when to refactor.

**(D15) "Constituting Constructor Parameters" are like fields**    We discuss calls from `ElbowHandle` to a `LineConnection` received from the superclass `AbstractHandle` via methods `AbstractHandle.owner()` as a `Figure` and `ElbowHandle.ownerConnection()` as a `LineConnection`.

In our discussion in the previous paragraphs we suggested that classes should represent the covariant type expected for their fields (when initialized in the constructor) by the type of the constructor parameter. Consequently, we should treat the constructor parameters like fields in the context of the "Law of Demeter". Let us capture the implicit assumption that constructor parameters are indeed always used for field initialization in the concept of "Constituting Constructor Parameter". In fact all constructor parameters in JHotDraw 5.1. are used for field initialization. Nevertheless there might be other uses of constructor parameters like in "Copy Constructors"[44] or constructor parameters that describe the modalities of creation.[45]

Since we argued that the type of "Constituting Constructor Parameter" should in general be considered to be like types of fields, we omit an discussion of A1) as well as A2). A3) We decide to consider all accesses to objects of type of an "Constituting Constructor Parameter" to be false positives.

---

[44] See for example the constructors `Rectangle(Rectangle r)` or `Point(Point p)` in java.awt.

[45] The signature of the constructor `Color(ColorSpace cspace, float components[], float alpha)` in java.awt might give the impression, that the color space is used only temporarily, especially when compared with other constructors like `Color(int r, int g, int b)`. If this were true, the parameter would not be a "Constituting Constructor Parameters" and the suggested adaptation would not apply. In fact the constructor does initialize a field representing the color space. The example should still be suggestive enough to illustrate the possibility of using parameters for other ends than field initialization.

I1) The design idea we refer to is "Constituting Constructor Parameters", i.e. constructor parameters used to build up the created object. I2) The definition of the "Law" should consider these constructor parameters to be like fields. I3) The design idea extends over all constructor parameters in JHotDraw 5.1, so that we don't need any restriction here. I4) Our confidence that the adaptation is justified in general is very high, as long as the software under consideration uses constructor parameters only to constitute new objects. I5) Considering accesses to types of "Constituting Constructor Parameter" to be false positives reduces the number of potential violations by 86.

**(D16) Refactoring: Specialize Constructor Parameter Type.**   In table 2 we mentioned that `FontSizeHandle` has an owner of type `TextFigure`. The owner is twice accessed by the expression "(TextFigure) owner()". Its constructor has the signature `FontSizeHandle(Figure owner, Locator l)` and thus does not ensure that the owner passed to the constructor has the required type. However, it is possible to change the signature to use the type `TextFigure`. With the suggested adaptation in the previous paragraph, this change removes as well potential violations.

We look at calls from `FontSizeHandle` to a `TextFigure` received from the superclass `AbstractHandle` as a `Figure`. Since the case is by the simple change of expressing the expected type in the signature of the constructor transformed into the situation discussed in (D15), we omit the discussion of A1) as well as A2) and A3) state that it is useful to consider the case to be a true positive, that can be easily resolved by changing the signature of the constructor.

Specializing a constructor parameter (always of type `Figure` to an appropriate specialized type) in the classes `ChangeConnectionEndHandle`, `ChangeConnectionStartHandle`, `ConnectedTextTool`, `FontSizeHandle`, `PolyLineConnector` resolves in total 19 potential violations.

### A.7  Observations in `DrawApplication`

Listing 9 shows one "Initialization Method" and three (plus five abbreviated) "Factory Methods" of the class `DrawApplication` plus the "Callback Method" `selectionChanged(DrawingView)`. The "Initialization Method" `createMenus(MenuBar)` populates on the start of the application the menu bar with five menus. The "Factory Method" for the attributes menu calls the "Initialization Method" `createFontMenu()` that populates a submenu with a list of all available fonts. The "Initialization Method" `createContents(StandardDrawingView)` creates within the drawing view a `ScrollPane`, i.e., an area that can be scrolled. In the "Callback Method" `selectionChanged(DrawingView)` for two of the menus in the menu bar the method `checkEnabled()` is called. These menus contain menu items for commands that are only available when figures are selected.

The presented methods of `DrawApplication` still contain seven potential violations that were not covered by the discussions in the previous sections. The code extract in table 9 is tailored for the discussion of the two calls to `checkEnabled()` on the menus retrieved from the menu bar via calls to `getMenu(int)` in the method `selectionChanged(DrawingView)`. These four calls are discussed in (D17). The access to the "field" `length` of the local variable fonts, an array of `String`, is discussed in (D18). Finally the calls to the method setUnitIncrement(int) on objects retrieved from a newly created ScrollPane in createContents(StandardDrawingView) is discussed in (D19).

**(D17) Refactoring: Keep specialized references.**　There are four potential violations in the method selectionChanged(DrawingView) in the class DrawApplication: The two calls to the method checkEnabled() in the class CommandMenu and the two calls to the method getMenu(int) in the class MenuBar. The two menus are instances of CommandMenu like the menu created in createFontMenu(). Unfortunately this type information is immediately given up. Furthermore the association from the class DrawApplication to its MenuBar is realized in the superclass Frame and thus as well not expressed by a field in DrawApplication.

We discuss calls from DrawApplication to a CommandMenu received from a MenuBar. The calls from DrawApplication to a MenuBar received from the superclass Frame will become obsolete as consequence of the result of this discussion.

A1) The potential violation creates coupling of low intensity to a stable class. The overall understandability of the menus being created and later on updated on selection change is not difficult. Reconstructing the relations might nevertheless require some scrolling (the extract shows only ~46 of 749 lines). The correctness of the constants EDIT_MENU and ALIGNMENT_MENU depends on the order in createMenus(MenuBar) and the convention to start the index at zero. The inner structure of the menu is known by the DrawApplication, but the class is as well the creator of the menu, so that the inner structure was in this aspect anyway not encapsulated. There is no problem in terms of coupling and missing encapsulation of the knowledge about the menu structure. Having the knowledge about the index of the menu items of interest at two locations—namely the definition of the constants and implicitly in the order of the creation—is nevertheless fragile.

A2) Since MenuBar is an external class, neither pushing nor lifting is possible. Nevertheless the code can be restructured along the idea of "Covariant Redundant Field" as presented in (D13). We initialize in the method createMenus(MenuBar) references editMenu and alignmentMenu of type CommandMenu, so that we do not need the casts in selectionChanged(DrawingView) anymore. The fragile indexing and the two integer constants are gone as well and we do not need to access the menu bar. We do need to change the result type of the "Factory Methods" to CommandMenu, which is possible since the created menus are indeed CommandMenus. While the two typical refactorings are not applicable, the potential violations can easily be solved by keeping specialized references. A3) The potential violations are more like a false positive, but very useful ones, that reveal other problems that can be successfully refactored away, resolving four potential violations.

**(D18) "Array Length"**　The method createFontMenu() in table 9 contains the expression "fonts.length" where fonts is an array of String. This looks like a field access. This "field" length of arrays is accessed eight times in JHotDraw 5.1. Although the access to length is syntactically like a field access, it is rather an expression of its own kind. This "field" can not be written and it does not belong to any class. As a consequence of the latter observation, no access of the field length would ever be allowed by the "Law".

**Did JHotDraw Respect the Law of Good Style?**

```
1   public class DrawApplication extends Frame implements DrawingEditor, PaletteListener {
2       public static final int EDIT_MENU = 1;
3       public static final int ALIGNMENT_MENU = 2;
4
5       protected void createMenus(MenuBar mb) {
6           mb.add(createFileMenu());
7           mb.add(createEditMenu());
8           mb.add(createAlignmentMenu());
9           mb.add(createAttributesMenu());
10          mb.add(createDebugMenu());
11      }
12
13      protected Menu createFileMenu() { /* [...] */ }
14      protected Menu createEditMenu() { /* [...] */ }
15      protected Menu createAlignmentMenu() { /* [...] */ }
16      protected Menu createAttributesMenu() { /* [ ... createFontMenu() ... ] */ }
17      protected Menu createDebugMenu() { /* [...] */ }
18
19      protected Menu createFontMenu() {
20          CommandMenu menu = new CommandMenu("Font");
21          String fonts[] = Toolkit.getDefaultToolkit().getFontList();
22          for (int i = 0; i < fonts.length; i++)
23              menu.add(new ChangeAttributeCommand(fonts[i], /* [...] */ ));
24          return menu;
25      }
26
27      protected Component createContents(StandardDrawingView view) {
28          ScrollPane sp = new ScrollPane();
29          Adjustable vadjust = sp.getVAdjustable();
30          Adjustable hadjust = sp.getHAdjustable();
31          hadjust.setUnitIncrement(16);
32          vadjust.setUnitIncrement(16);
33          sp.add(view);
34          return sp;
35      }
36
37      public void selectionChanged(DrawingView view) {
38          MenuBar mb = getMenuBar();
39          CommandMenu editMenu = (CommandMenu)mb.getMenu(EDIT_MENU);
40          editMenu.checkEnabled();
41          CommandMenu alignmentMenu = (CommandMenu)mb.getMenu(ALIGNMENT_MENU);
42          alignmentMenu.checkEnabled();
43      }
44  }
```

The calls in createFontMenu() to getDefaultToolkit() and createFontList() are false positives since (D4) and (D10). The access to the "field" length of the array fonts[] appears only to be a violation, because this "field" is in no class (D18). As discussed in (D19) the two calls to setUnitIncrement(int) in the method createContents(StandardDrawingView) are strong breaches of the encapsulation of the class ScrollPane. Unfortunately the relevant code is not under our control so that we have to resign and to consider the calls as practically false positives. Since the potential violations in selectionChanged(DrawingView) can easily be solved (D17) we consider them as true positives.

We discuss calls from `DrawApplication` to an array of `String` received from a `Toolkit`. A1) The field is a language construct and thus most stable. Understanding `length` is as easy as understanding the idea of an array in general and no hidden information is revealed by the usage of the field. A2) If `Toolkit` was not external, one could consider lifting the access to the field `length` to a method `getFontListLength()`. This would be very uncommon and useless in cases, where we need the full array anyway. Even if we could push the responsibility of menu population to `Toolkit` this responsibility would have nothing to do with responsibility of that class. A3) Since the use of the "field" is completely unproblematic and the suggested (technically impossible) refactorings are absurd, the potential violation is without any doubt a false positive.

I1) Lets refer to the "field" `length` in arrays by the name "Array Length" I2) and consider "Array Length" as "everybody's friend member". I3) The "idea" just refers to this "field". We intended to discuss the potential violations without any reference to the implementation that we used. This case might be an exception in so far as it is not clear, what kind of animal `length` is. In our representation of the source it is modeled as a field. I4) There is no doubt about this adaptation and I5) it reduces the number of potential violations by eight.

**(D19) "Designated Accessor in External Code"** The method `createContents(StandardDrawingView)` in table 9 contains two calls to the method `setUnitIncrement(int)` of the type `Adjustable`. The instances are retrieved from a newly instantiated `ScrollPane`. Since there is no other relation of the method to the type `Adjustable`, the calls to `setUnitIncrement(int)` are potential violations.

We discuss calls from `DrawApplication` to `Adjustables` received from a `ScrollPane`. A1) The coupling toward the stable class has low intensity. The overall construction is rather uncommon. The objects `hadjust` and `vadjust` must still have a connection to the object `pane` for the method calls to make sense. The encapsulation of `ScrollPane` is thus strongly violated. We do not only learn something about the internal structure of `ScrollPane`, but we even get access to the internal state through these objects. This design is in direct opposition to the "Law of Demeter". A2) Since `ScrollPane` is an external class neither lifting nor pushing is possible. A3) This potential violation is thus a true positive, but a useless one, since the relevant code is not under our control. We therefore suggest to resign and to consider this potential violation with deep regret to be a practically false positive.

I1) The code of the JDK gives us no chance to call `setUnitIncrement(int)` than to retrieve objects through "Designated Accessors in External Code". I2) Since we have no other choice, we suggest to consider the result type of such a method to be a "friend" type to any method, who calls it. I3) The methods forcing us into this situation can be enumerated. I4) Since we consider only methods that give us no other choice, the adaptation is a necessary consequence and we are confident that the respective adaptation is required. I5) In this way five similar cases are "resolved", i.e. removed from our consideration.

There might be other perspectives on the given example. The original developers might have seen the objects retrieved from the `ScrollPane` to be figuratively "Remote Control Objects" or part of a "Configurable View Model", so that these objects would have been meant to control the pane. The original developers might have had in mind to restrict the use of these objects, e.g. never passing them around or never allowing any aliases. Under such additional constraints even the obvious violation of the encapsulation might be tolerable.

## A.8  State of the overall discussion

Till here only 76 or 6.3 % of the potential violations are not yet discussed. We found in total in (D3), (D16) and (D17) $6 + 19 + 4 = 29$ true positives. For the $1215 - 29 - 76 = 1110$ false positives suggestions were given how to adapt the original definition of the "Law" based on design ideas to exclude the false positives from the heuristic. Till now all adaptations were independent of the concrete software under discussion. Only for the design ideas "Data Class", "Singleton", "Factory Method" and "Designated Accessor in External Code" it was suggested to define their extension by enumerating the names of the respective classes or methods. Naming these few program elements is little effort and much more reliable than the introduction of new heuristics that might be imprecise themselves. The design idea "Constituting Constructor Parameter" extends over all constructor parameters in JHotDraw 5.1. As it tries to capture some extra intension, it might become necessary for other software to narrow its extension.

The effort for reviewing 76 potential violations during day to day software development is more reasonable than to review 1215, but still too tedious to be done regularly. From the perspective of the exploratory case study stopping here is anyway no option. The goal is to cover all remaining potential violations as well. Since we preferred till now discussions that gave us the opportunity to exclude a higher number of false positives based on rather general design ideas, we can expect to find now the opposite: Discussions that depend more on the specifics of JHotDraw 5.1 and that might not be as effective in the elimination of false positives as before. We may as well expect to discover design ideas that give us more specific information about the software.

## A.9  Observations in SelectAreaTracker

The class `SelectAreaTracker` is responsible for drawing a rectangle ("rubber band") on the drawing view in response to mouse actions and to finally select the figures enclosed in the "rubber band". Listing 10 shows the methods for drawing this "rubber band" and for selecting the figures.

In the methods `drawXORRect(Rectangle)` and `selectGroup(boolean)` the suggested adaptations in (D5) and (D15) let us already consider six potential violations to be false positives. The remaining potential violation in `drawXORRect(Rectangle)` are calls to three different methods on an object of type `Graphics` retrieved from the view. The two remaining potential violations in `selectGroup(boolean)` are a call to the method `figureReverse()` on the drawing of the view accessed via an accessor defined in the class `AbstractTool`[46]

---

[46] This potential violation will be discussed as well implicitly in (D24).

■ **Listing 10**  Extract from the class SelectAreaTracker in JHotDraw 5.1 (∼44 of 77 lines).

```
1   public class SelectAreaTracker extends AbstractTool {
2       private Rectangle fSelectGroup;
3
4       public SelectAreaTracker(DrawingView view) {
5           super(view);
6       }
7
8       // [Create and update the "rubber band". Finally select enclosed figures.]
9       public void mouseDown(MouseEvent e, int x, int y) { }
10      public void mouseDrag(MouseEvent e, int x, int y) {}
11      public void mouseUp(MouseEvent e, int x, int y) { }
12
13      private void rubberBand(int x1, int y1, int x2, int y2) {
14          fSelectGroup = new Rectangle(new Point(x1, y1));
15          fSelectGroup.add(new Point(x2, y2));
16          drawXORRect(fSelectGroup);
17      }
18
19      private void eraseRubberBand() {
20          drawXORRect(fSelectGroup);
21      }
22
23      private void drawXORRect(Rectangle r) {
24          Graphics g = view().getGraphics();
25          g.setXORMode(view().getBackground());
26          g.setColor(Color.black);
27          g.drawRect(r.x, r.y, r.width, r.height);
28      }
29
30      private void selectGroup(boolean toggle) {
31          FigureEnumeration k = drawing().figuresReverse();
32          while (k.hasMoreElements()) {
33              Figure figure = k.nextFigure();
34              Rectangle r2 = figure.displayBox();
35              if (fSelectGroup.contains(r2.x, r2.y)
36                      && fSelectGroup.contains(r2.x+r2.width, r2.y+r2.height)) {
37                  if (toggle)
38                      view().toggleSelection(figure);
39                  else
40                      view().addToSelection(figure);
41              }
42          }
43      }
44  }
```

In (D15) we had already decided to accept constructor parameters like DrawingView to be like fields and in (D5) we had decided to tolerate calls to collection methods. The only remaining potential violations are the calls on the local variable g of type Graphics in the method drawXORRect(Rectangle) and the calls to Drawing.figuresReverse() and to Figure.displayBox() in the method selectGroup(boolean). Our discussion in (D20) leads to the suggestion to push the responsibility to draw an rectangle in "XOR-mode", the "rubber band", to the view. The same is true for selecting the figures enclosed in the "rubber band".

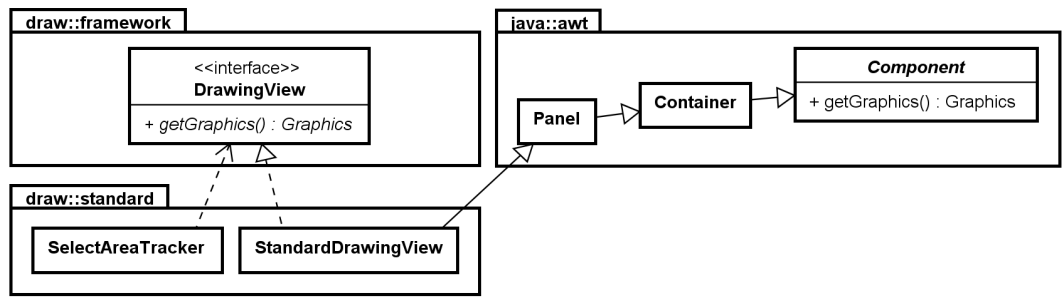

■ **Figure 9**   SelectAreaTracker gets the Graphics objects from far away.

and a call to the method `displayBox()` on all figures retrieved from the drawing. We discuss these cases in (D20).

**(D20) Refactoring: Push drawing the "rubber band" back.**   `SelectAreaTracker` contains in the method `drawXORRext(Rectangle)` in table 10 three method calls to methods of the class `Graphics`. The instance of `Graphics` is retrieved from the `DrawingView`. There are furthermore the calls to `Drawing.figuresReverse()` and to `Figure.displayBox()` in the method `selectGroup(boolean)`. Our discussion will mainly focus on the violations in `drawXORRect(Rectangle)`.

We discuss calls from `SelectAreaTracker` to a `Graphics` object received from a method declared in `DrawingView` which has no implementation in JHotDraw 5.1 but is implemented in java.awt.Component. More precisely, `StandardDrawingView` implements `DrawingView` but has no definition of the method `getGraphics()`. The implementation is inherited from its superclasses in java.awt: StandardDrawingView extends java.awt.Panel which extends Container which finally extends Component, see figure 9.

A1) The calls couple `SelectAreaTracker` to the presumably stable class `Graphics`. Since three methods are accessed, the coupling is more intense than we mostly found before. The methods draws a rectangle on the view in "XOR-Mode". Latest after consulting the documentation of this method this should be understandable as well to a new developer. A bit more experienced developer might wonder, whether it is OK, to just start drawing on the `Graphics` object. Often drawing is structured and controlled by views, e. g. by setting up colors, clipping areas and so on. Deciding, when and how to draw is a typical responsibility of a view. Consequently the encapsulation of the `DrawingView` might indeed be considered to be broken. In addition the association to `Graphics` is deeply buried and quite artificially unearthed through the interface. Taking the slightly more intense coupling, the moderate understandability and the doubts about the allocation of the responsibility into account, it would be nice to find a better design.

A2) Lifting the three additional methods into the interface `DrawingView` is no option. These methods are not used by any other class and thus lifting would fix no other violations and decouple no other classes. Pushing the responsibility of the whole method to `DrawingView` on the contrary is a good choice. The getGraphics() method could then be deleted from the interface. We have to add one (`drawXORRect(Rectangle)`) or two (`rubberBand(int, int, int, int)` and `eraseRubberBand()`) methods in interface and imple-mentation. The responsibility is consistent with the overall responsibilities of the view, namely drawing and selection management. Although the new methods would currently only used by the SelectAreaTracker one could imagine other uses, e.g. highlighting the area that is presented in a second view, maybe magnified. While the lifting refactoring is no option the pushing refactoring is a perfect fit.

A3) Since the current situation asks for improvement and the pushing refactoring offers it, these three potential violations are true positives. Since the selectGroup(boolean) fits as well nicely to the other responsibilites of the DrawingView, we suggest to push it as well. So that in total five violations are resolved.

### A.10  Observations in AlignCommand.execute()

The class AlignCommand is responsible for aligning the currently selected figures to the left, right, up, down or to center them horizontally or vertically. Listing 11 shows the code responsible for aligning figures to the left. The method execute() contains the actual operation on the figures.

The only remaining potential violations in the method execute() are two calls to the method displayBox() and one call to the method moveBy(int, int) and will be discussed in (D21). A very similar case will be discussed in (D22) without explicit code.

**(D21) "DrawingView Manages the Selection"**    The method execute() implemented in the class AlignCommand as presented in table 11 calls displayBox() and moveBy(int, int) of the type Figure to align the selected figures. These instances of Figure are retrieved from a FigureEnumeration retrieved from the DrawingView through the method selectionElements(). The method execute() has no direct relationship to the type Figure.

We discuss calls from AlignCommand to Figures received from a DrawingView. A1) Two methods of the stable interface Figure are called to align the figures after the selected figures had been received from the view. This is easy to understand. Managing the selection is one of the main responsibilities of the view. Receiving an enumeration of the selected figures is the intended way of sharing this information. Thus, encapsulation of the view is not violated. The current design is good. A2) If we wanted to lift the methods to the DrawingView we would need at least three methods, since we need to access the display box of the first selected figure and then the remaining selected figures and move those accordingly. A responsibility that we could push would be the whole alignment process. This would create one additional method that no other class would use. The responsibility also does not fit in the DrawingView since there is no manipulation of figures in the DrawingView. We could try to push it into Drawing but there is no handling of the selection in that class. Other commands keep as well manipulation logic and typically use the selection information from the view to manipulate the drawing. The current distribution of responsibilities is consistent with the rest of the application. A3) Changing the given design would be against the intention and responsibility distribution of these central classes. The potential violation should be considered to be a false positive.

I1) "DrawingView Manages the Selection" of Figures, but other classes are allowed to work on these selected figures. Having the information about the selection nowhere

**Did JHotDraw Respect the Law of Good Style?**

```java
1   public class AlignCommand extends Command {
2
3       private DrawingView fView;
4       private int fOp;
5
6       public final static int LEFTS = 0;
7       // [ ... five more directions ... ]
8
9       public void execute() {
10          FigureEnumeration selection = fView.selectionElements();
11          Figure anchorFigure = selection.nextFigure();
12          Rectangle r = anchorFigure.displayBox();
13
14          while (selection.hasMoreElements()) {
15              Figure f = selection.nextFigure();
16              Rectangle rr = f.displayBox();
17              switch (fOp) {
18              case LEFTS:
19                  f.moveBy(r.x-rr.x, 0);
20                  break;
21              // [ ... five more directions ... ]
22              }
23          }
24          fView.checkDamage();
25      }
26  }
```

Besides the already in (D5) discussed accesses to collection methods, there remain the calls to `displayBox()` and `moveBy(int, int)` of the type `Figure` in the method `execute()`. These point us to a central design decision in JHotDraw 5.1, namely that the view is responsible for maintaining the list of selected figures and that all other objects that want to know this selection, need to ask the view for it. Since this is a reasonable design, we suggest in (D21) to consider accesses to `Figure` to be acceptable in all methods that ask the view for the current selection.

duplicated helps to keep the logic consistent. The price for this consistency are 15 accesses to `Figure`s from methods, for which `Figure` is not a "friend". The idea that selections are held by `DrawingView`s is essential to JHotDraw and is depicted in the "class diagram showing the most important classes and their relationships".[47] I2) For all methods calling `selectionElements()` and similar methods on the interface `DrawingView` the type `Figure` should be considered "friend". I3) The methods to access the selection in the `DrawingView` follow the naming pattern `selection*(..)`. I4) Since the centralization of the selection handling is well motivated and results in the indirect access to `Figure` the given argumentation holds not only for `AlignCommand.execute()` but for all methods using the selection. I5) The adaptation removes 20 potential violations.

---

[47] See documentation/documentation.html in JHotDraw 5.1.

**(D22) "Clipboard Contains Figures"** There is a very similar case for which we omitted the listing. The method `execute()` in the class `PasteCommand` retrieves from the "Singleton" `Clipboard` an instance of the class `FigureSelection` containing `Figures`. The method `getData(String)` is called on the `FigureSelection`.

We discuss calls from `PasteCommand` to a `FigureSelection` received from a `Clipboard`. A1) The calls are overall unproblematic very similar to (D21). The `Clipboard` contains a collection of figures to be transfered. The access to `Clipboard` content is intentionally untyped. A2) Pushing or lifting methods to the `Clipboard` that are related to other responsibilities than transferring figures is no option. A3) The comments explain that the `Clipboard` class is a temporary substitute for a clipboard from the framework. Otherwise adding more type information to the code would be an option. Following this intention, the given potential violations are false positives.

I1) If the "Clipboard Contains Figures" I2) methods receiving the content of the clipboard should consider `Figure` and `FigureSelection` to be a "friend" type. I3) The `Clipboard` class can be identified by name and package. I4) As long as the `Clipboard` is accepted as an untyped means to transfer figures, the given violations are a natural consequence. I5) This adaptation resolves two potential violations.

## A.11 Observations in ActionTool

The class `ActionTool` is an abstract root class for any tool that just executes one action on a mouse click. In JHotDraw 5.1 there is only one tool like this, namely the `BorderTool` that adds a border to the figure receiving the mouse click. The 20 lines in table 12 show the complete code of the class `ActionTool`, only comments and the package and import declarations were removed. The method `mouseDown(MouseEvent, int, int)` finds the figure at the position of the click and invokes the abstract action on it. The method `mouseUp(MouseEvent, int, int)` just sends a notification that the work of the tool is done.

The constructor of the `ActionTool` receives a reference to a `DrawingView` and passes it on to its superclass `AbstractTool`. The `AbstractTool` provides not only access to the `DrawingView` through the accessor `view()` but as well to the `DrawingEditor` through the accessor `editor()` and to the `Drawing` through the accessor `drawing()`. All three accessors are used in `ActionTool`. `AbstractTool` can provide this access, because the `DrawingView` provides access to the `DrawingEditor` and to the `Drawing`. Since (D15) the call to `addToSelection(Figure)` on the result of `view()` is no longer considered a violation. The call to `toolDone()` on the result of `editor()` is discussed in (D23) and the call to `findFigure(int, int)` on the result of `drawing()` in (D24). Similar to `DrawingView` `DrawingEditor` also provides access to other classes inviting in the same way violations of the "Law of Demeter". This will be discussed in (D25).

**(D23) Refactoring: Push notification up to superclass.** The method `mouseUp(MouseEvent, int, int)` in the class `ActionTool` as presented in table 12 retrieves a reference to the `DrawingEditor` through an accessor `editor()` of its superclass and calls the method `toolDone()` on it. Consulting all the uses of the accessor one finds that immediately calling `toolDone()` is the only thing the reference to the

**Did JHotDraw Respect the Law of Good Style?**

■ **Listing 12**   The class ActionTool in JHotDraw 5.1 (20 of 43 lines)

```
 1  public abstract class ActionTool extends AbstractTool {
 2
 3      public ActionTool(DrawingView itsView) {
 4          super(itsView);
 5      }
 6
 7      public void mouseDown(MouseEvent e, int x, int y) {
 8          Figure target = drawing().findFigure(x, y);
 9          if (target != null) {
10              view().addToSelection(target);
11              action(target);
12          }
13      }
14
15      public void mouseUp(MouseEvent e, int x, int y) {
16          editor().toolDone();
17      }
18
19      public abstract void action(Figure figure);
20  }
```

ActionTool inherits from AbstractTool methods to access the DrawingView and through this view the Drawing and the DrawingEditor. Our discussion in (D15) accepts the access to DrawingView because of the type of the constructor parameter. The DrawingView gives tools, handles and commands direct access to the Drawing. We consider this design decision to be justifiable and suggest therefore in (D24) to allow every object that "is friend with" the view to access the drawing as well. The dependency to DrawingEditor turns out to be unnecessary since all tools use it for the same purpose of notifying about completion. As discussed in (D23) this responsibility can therefore easily be pushed upward into the AbstractTool.

DrawingEditor is ever used for in any tool. Six tools execute editor().toolDone() either in mouseUp(MouseEvent, int, int) or mouseDown(MouseEvent, int, int).

We discuss calls from ActionTool to a DrawingEditor received from the superclass AbstractTool. A1) Since DrawingEditor is a general interface and ActionTool a specifc class, the former can be expected to be at least as stable as the latter. Just one method is called, so that the intensity is as well low. Hence we have coupling of low intensity in the direction of stability, which is no problem. The method name does not tell us what the editor will be doing, but that the editor is notified about the completion of whatever the tool had to do. This is easy to understand. Whether or not the encapsulation of the AbstractTool or the DrawingView is violated remains unclear. It would be surprising if the editor would be a part of the tool or the view. This unclear relation to the editor is the critical point of the given potential violation. A2) Since the method under consideration contains just the call to toolDone() on the result of editor() the only responsibility that could be pushed is this call. This is the same as lifting the call forward. One would add a method toolDone() to the class AbstractTool. Since these calls are the only reasons for the method editor() it can in return be removed. The size of the interface stays the same. Violations in five classes can be removed and only AbstractTool knows the editor. The new method represents a meaningful step in the activity cycle of a tool and thus fits nicely into the class. The refactoring is a good choice. A3) Since the refactorings concentrates the unclear relation to the editor in one class in a method that neatly fits in, the potential violation should be considered to be a true positive. The refactoring resolves six identical cases.

**(D24) "DrawingView Provides Access to its Drawing"**  In the class `ActionTool` presented in table 12 we find the method `mouseDown(MouseEvent, int, int)`. It calls the method `findFigure(int, int)` on an instance of type `Drawing`. The reference to this object is retrieved through the accessor `drawing()` of the `AbstractTool`. Consulting all the uses of this accessor one finds in contrast to (D23) different uses of the `Drawing`: finding figures with and without descending into groups, finding `ConnectionFigures`, removing a figure, or replacing a figure.

We discuss calls from the class `ActionTool` to a `Drawing` received from the superclass `AbstractTool` and indirectly from `DrawingView`. A1) The coupling has low intensity and is directed to a more stable class. The functionality of the methods is easy to understand. The method tries to find a figure at the coordinates of the moue click. If successful, it adds the figure to the selection and executes an action on it. Whether or not the encapsulation of the `AbstractTool` or the `DrawingView` is violated is again unclear. In contrast to the relation between tool, view and editor in (D23) it is plausible to consider the drawing to be a part of the view and the `AbstractTool` would just provide a convenient access for its subclasses. The encapsulation of the view would thus have been broken, but at least the relation between the classes is clearer than in (D23). Whether the current design is to be considered bad depends on how we think about this breach of encapsulation. In terms of coupling and understandability it is fine.

A2) Lifting the method `findFigure(int, int)` to `AbstractTool` or `DrawingView` would increase the size of the interface by one. We could decouple this way three tool classes from `Drawing`. If we wanted to decouple as well the other classes, we would need to lift four more methods. These five methods fit better to the responsibilities of the `Drawing` than to the alternative classes, so that we would not recommend lifting. Pushing the single statement would be the same as the lifting refactoring. Pushing more would create a method in `AbstractTool` that would not even be used by any other tool. Pushing to `DrawingView` would not be consistent with the responsibilities of the view. A3) None of the refactorings is convincing. To discuss whether the broken encapsulation is critical, we review as well some other classes. We find that to really decouple the tools from the `Drawing`, we would mix client/subclass specific methods into the `AbstractTool` or `DrawingView`. When we consult the command classes we find that they operate on the selection in the `DrawingView` view as well as on the state in the `Drawing`. Therefore giving direct access to the `Drawing` is preferable and we consider the potential violation to be false positive.

I1) Let us capture this design idea as "`DrawingView` Provides Access to its Drawing" and I2) consider everywhere, where `DrawingView` is considered a "friend" as well `Drawing` to be a "friend". In terms of the analogy, we could call the `Drawing` a "best buddy" of the `DrawingView`, so that you do not get around considering the former your "friend" when you want to consider the latter to be your "friend". I3) The extension of the idea is just given by naming the two classes. I4) Since the suggested adaptation is a result of

**Did JHotDraw Respect the Law of Good Style?**

■ **Listing 13**   The interface DrawingEditor in JHotDraw 5.1 (17 of 59 lines)

```
1   /**
2    * DrawingEditor defines the interface for coordinating the different objects that participate in
3    * a drawing editor. [...] DrawingEditor is the mediator. It decouples the participants of a
4    * drawing editor. [...]
5    */
6   public interface DrawingEditor {
7
8       DrawingView view();
9       Drawing drawing();
10      Tool tool();
11
12      void toolDone();
13      void selectionChanged(DrawingView view);
14      void showStatus(String string);
15
16  }
```

Although the comment claims that the DrawingEditor is a "Mediator" that *decouples* its "participants" (view, drawing, tools), it *exposes* them via the methods view(), drawing(), tool(). Consequently methods referring to the DrawingEditor may depend also on Drawing, DrawingView or Tool, potentially resulting in "Law of Demeter" violations. Nevertheless the code expresses this questionable design intention so clearly that we suggest in (D25) to ignore the resulting violations. They require a deeper discussion of the design. We found some technical debt that can not easily be repaid.

the distribution of responsibilities between the classes, it can be generalized to other potential violations with confidence. I5) There are in total 28 similar false positives that we remove with this adaptation of the "Law".

**(D25) "DrawingEditor Exposes Colleagues".**    At the beginning of this section we explained that AbstractTool is able to provide access to the DrawingEditor and the Drawing because DrawingView provides access to instances of these two classes. As table 13 shows, DrawingEditor is similar. It provides access to the DrawingView, the Drawing and the current Tool. Consequently any class that has a reference to a view can access as well editor, drawing and tool. Similarly any class that has a reference to an editor can access as well view, drawing and tool.

The comments for the interface DrawingEditor claim that "DrawingEditor is [a] mediator [and it] decouples the participants of a drawing editor." See table 13. Decoupling one object from another means that one object has neither a direct reference to the other object nor that it knows the type of the other object. In the "Mediator" pattern as described in [13, pp. 273-282] the "Colleagues" (not "Participants" as in the comment) know only the "Abstract Mediator" type and use it to send commands or notifications to the mediator object. In the "Concrete Mediator" type is defined how the mediator object reacts. Typically this involves calling methods of some of the "Colleagues". The three methods toolDone(), selectionChanged(DrawingView) and showStatus(String) fit well into this scheme. The three methods view(), drawing() and tool() in contrast corrupt the intention of the "Mediator" pattern, since they expose the types of the "Colleagues" and provide access to them. Since these accessors may give access to types even when they are not "friend" types in the sense of the "Law of Demeter", they invite violations of the "Law".

Luckily this invitation to corrupt the "Mediator" was not widely accepted by the developers. The accessor view() is only used in the implementations of the interface DrawingEditor to access the view. The accessor drawing() is only used in two of these implementations to start or stop the animation and to pass the drawing to a new window. Both declarations could be deleted without further consequence, since in these classes already the implementation of the methods is known not only the declarations in the interface. The accessor tool() is only used in StandardDrawingView to send MouseEvents to the current tool. These method calls are the only potential violations resulting from the DrawingEditor interface.

We discuss calls from StandardDrawingView to a Tool received from a DrawingEditor. A1) The calls create coupling to a stable interface. Each call creates just coupling to one method, but together five methods are used. What they are used for, is easy to understand: The calls send slightly preprocessed mouse and keyboard events to the tool and update the view afterward. If the tool is a part of the DrawingEditor, the editor breaks its encapsulation by giving access to the tool. But, is the tool actually a part of the editor? The answer to this question is not obvious. The isolated violations are rather unproblematic, the overall relation between editor, view and tool is unclear. A2) Lifting the access to Tool to the DrawingEditor would mean to add five methods for receiving mouse or keyboard events to this central interface, just to decouple StandardDrawingView. This is no option. Since the five calling methods are different, pushing would as well mean creating five new methods in the DrawingEditor interface. This is again no option. The situation can be solved, but it requires a more complex redesign. One option could be to move the association to the Tool from the DrawingEditor to the DrawingView. The DrawingEditor would then need to change the current tool in the DrawingView anytime a new tool is selected or a tool is done. Meanwhile the view could us the tool without any indirection. While lifting and pushing are no option, moving the association could resolve the potential violations. A3) The overall relation between the classes is not that clear and the potential violations could be resolved by moving the association to Tool from DrawingEditor to DrawingView. This would be a big change to resolve "just" a "Law of Demeter" violation and requires some design evaluation. The overall design intention is here less clear than in (D24) and the interface in table 13 is in itself contradictory, as far as it claims to be a "Mediator" but exposes the "Colleagues" through its three accessor methods. We suggest to consider the potential violation as "false until a design review".

I1) We describe the current situation as "DrawingEditor Exposes Colleagues". I2) A method that is "friend" with DrawingEditor may be considered to be as well "friend" with Tool. I3) The two classes are identified just by their name. I4) For the time being we have enough confidence in this adaptation. After a thorough design discussion, the potential violations might be removed as discussed above. I5) Although the exposure of its "Colleagues" by the DrawingEditor could have had a major impact, there are actually only five potential violations. These calls to Tool in StandardDrawingView are covered by the given adaptation.

■ **Listing 14** Extract from the class `PertFigure` in JHotDraw 5.1 (31 of 296 lines)

```java
1   public class PertFigure extends CompositeFigure {
2
3       public PertFigure() {
4           initialize();
5       }
6
7       private int asInt(int figureIndex) {
8           NumberTextFigure t = (NumberTextFigure)figureAt(figureIndex);
9           return t.getValue();
10      }
11
12      private String taskName() {
13          TextFigure t = (TextFigure)figureAt(0);
14          return t.getText();
15      }
16
17      private void setInt(int figureIndex, int value) {
18          NumberTextFigure t = (NumberTextFigure)figureAt(figureIndex);
19          t.setValue(value);
20      }
21
22      private void initialize() {
23          // [ ... Initialization code here and for the figures below deleted ... ]
24          TextFigure name = new TextFigure();
25          add(name);
26          NumberTextFigure duration = new NumberTextFigure();
27          add(duration);
28          NumberTextFigure end = new NumberTextFigure();
29          add(end);
30      }
31
32  }
```

The `PertFigure` makes use of the ability of its superclass `CompositeFigure` to manage a collection of figures. In contrast to its superclass it requires these figures to be of specific type. Although it thus violates - strictly speaking - the contract of its superclass, we consider this design to be acceptable and suggest in (D26) to consider the access to `TextFigure` and `NumberTextFigure` in the class `PertFigure` to be as well acceptable.

## A.12 Observations in `PertFigure`

One of the sample applications delivered together with JHotDraw 5.1 allows to draw PERT diagrams [23]. The figures for these diagrams are implemented by the class `PertFigure` and `PertDependency`. As can be seen from table 14 a `PertFigure` is a `CompositeFigure` (see the `extends` declaration) built from one `TextFigure` and two `NumberTextFigures` (see the `initialize()` method). The methods `taskName()`, `asInt(int)` and `setInt(int, int)` give private access to the values stored in these subfigures.

The three potential violations are calls to these subfigures. Since the aggregation of subfigures is realized in `CompositeFigure` it only contains `Figures`. We discuss in (D26) how to account for the fact that PertFigure relies on the specific types. (D27) discusses an almost identical situation in BouncingDrawing.

**(D26) "PertFigure is Composite of Specific Figures"**  The private methods asInt(int) and setInt(int, int) in the class PertFigure as presented in table 14 call the methods getValue() or setValue() of the type NumberTextFigure. The type system just guarantees that the objects on which these methods are called are of type Figure, but the developers had enough confidence that their code guarantees in the context of the sample application, that the figures actually have the more specific type, so that they can safely cast them to NumberTextFigure. The private method taskName() in the class PertFigure calls the method getText() of the type TextFigure. Again the type system just guarantees that the object is of type Figure but a cast expresses the confidence that it actually is a TextFigure.

We discuss calls from PertFigure to a TextFigure and NumberTextFigures received from the superclass CompositeFigure. A1) The coupling is toward a more stable class and has low intensity. Understanding the methods is straightforward. The idea that a "Composite" CompositeFigure aggregates Figures needs of course be understood. Given that, there is nothing revealed about CompositeFigure that is not known anyway and thus encapsulation is pertained. There is no problem with the given code in terms of coupling, understandability and encapsulation. It needs to be mentioned that PertFigure is not substitutable for CompositeFigures since you could e.g. not add a randomly chosen figure to it. The class is thus only usable in the context of a specific application that respects these limitations. A2) Pushing or lifting functionality into the CompositeFigure just for the purpose of one specific subclass is no option. A refactoring of keeping a specialized reference for the specific fields like in (D17) is of course possible. But, since the specific objects are created inside the class in the method initialize() there is less need to make the type information explicit. A3) We therefore consider the current design to be acceptable and consider the three potential violations to be false positives.

I1) We give the design idea the name "PertFigure is Composite of Specific Figures" and I2) consider the two specific components TextFigure and NumberTextFigure to be "friends" of the PertFigure.[48] I3) The idea does extend to the given class. I4) Since the context for adaptation is very specific our confidence in the adaptation is high. I5) The potential violations that are covered by this adaptation are the three presented calls.

**(D27) "BouncingDrawing is composite of AnimationDecorators"**  BouncingDrawing is a subclass of StandardDrawing which is a subclass of CompositeFigure. Similar to PertFigure as discussed in (D26) it relies on the fact that it contains specific figures, here AnimationDecorators. BouncingDrawing wraps any new figure with an AnimationDecorator, when added with add(Figure) or replace(Figure, Figure) and removes the wrapper, when the figure is deleted.

---

[48] We might as well take the casts on the result of figureAt(int) as introducing the type as "friend" of the method, but this inference would only become interesting, if we had reason to use it in more places.

The method `remove(Figure)` in the class `BouncingDrawing` calls `peelDecoration()` from `AnimationDecorator`. The method `animationStep()` calls `animationStep()` from `AnimationDecorator` on all figures. Since the `BouncingDrawing` uses the ability of `CompositeFigure` to maintain a list of figures, the type system just ensures that it contains `Figures`. Nevertheless — as described above — all figures are wrapped in `AnimationDecorator`.

We discuss calls from `BouncingDrawing` to a `AnimationDecorator` received from the superclass `CompositeFigure` as `Figure`. A1) Since `BouncingDrawing` and `AnimationDecorator` seem to be written to collaborate together they are of similar stability. The intensity of the coupling is as well low. The developer needs to understand that every (first level) object has to be wrapped in `AnimationDecorator` and unwrapped, when removed. In addition he needs to understand that the superclass `CompositeFigure` is responsible for maintaining the list of component objects. Once these two facets are understood, the code is clear. There is not more about the `CompositeFigure` revealed, than what is known anyway, namely that it aggregates `Figures`. Thus, encapsulation is preserved. The class has unproblematic coupling and good understandability and encapsulation. A2) Pushing or lifting concerns of the animation to `CompositeFigure` is no option, since it would mix a concern into that class that does not belong there. The two strategies to preserve the covariant type information are as well not feasible here. To keep a "Covariant Redundant Field" one would need to maintain an additional list of the component objects. Wrapping casts in "Covariant Accessor Methods" is no option, since methods are called from other classes with no guarantee of the object being wrapped in `AnimationDecorator` or not being wrapped. A3) We consider these potential violations to be false positives. The accessed objects are meant to be parts of the composite. The static type information was traded for the reuse of the general composite mechanisms.

I1) Let us name this situation "`BouncingDrawing` is composite of `AnimationDecorator`" I2) and consider `AnimationDecorator` to be a "friend" of `BouncingDrawing`. I3) The extension of this idea is given as the classes `BouncingDrawing` and `AnimationDecorator`. I4) The adaptation can be trusted since it is specific and there is a clear reason, why the type information is not available. I5) The adaptation impacts the two mentioned potential violations.

The variations to the "Composite" pattern described in (D26) and (D27) are similar in that in both cases the composite aggregates "Components" of specific type. They are nevertheless very different. The `PertFigure` relies so strongly on the first three components being of their specific type, that it can not easily be used in other applications than the Pert-application. In contrast the `BouncingDrawing` still fulfills the contract of the `CompositeFigure`. Any `Figures` can be added and removed from the figure. The `BouncingDrawing` will wrap them into an `AnimationDecorator` when added and unwrap them, when removed.

## A.13  Observations in PertDependency.handleConnect(Figure, Figure)

The method `handleConnect(Figure, Figure)` in the class `PertDependency` as presented in table 15 is invoked after the connection between two PERT figures has been established to update further attributes of the connected figures. The method relies on the pa-

■ **Listing 15**   Extract from the class `PertDependency` in JHotDraw 5.1 (19 of 62 lines)

```
1  public class PertDependency extends LineConnection {
2
3      public void handleConnect(Figure start, Figure end) {
4          PertFigure source = (PertFigure)start;
5          PertFigure target = (PertFigure)end;
6          if (source.hasCycle(target)) {
7              setAttribute("FrameColor", Color.red);
8          } else {
9              target.addPreTask(source);
10             source.addPostTask(target);
11             source.notifyPostTasks();
12         }
13     }
14
15     public boolean canConnect(Figure start, Figure end) {
16         return (start instanceof PertFigure && end instanceof PertFigure);
17     }
18
19 }
```

When the class `PertDependency` overrides the method `LineConnection.handleConnect(Figure, Figure)` it can not change the type of the method's parameter. Nevertheless other classes ensure using `canConnect(Figure, Figure)` that the actual arguments are always of the specific type `PertFigure`. The discussion in (D28) therefore suggests that not only `Figure` but as well `PertFigure` should be considered a "friend" type of the method.

rameters being of type `PertFigure` although the declared parameter type is only `Figure`. The parameter types have to be the same as the parameter types of the overridden method in `LineConnection`. There are four potential violations in the method. These are calls to methods declared in `PertFigure` but not yet in `Figure`. We will discuss these in (D28). Another kind of potential violations resulting from missing type information in parameters will be discussed in (D29).

**(D28) "Downcast Parameter"**   The method `handleConnect(Figure, Figure)` in the class `PertDependency` as presented in table 15 calls the methods `hasCycle(Figure)`, `addPreTask(PertFigure)`, `addPostTask(PertFigure)` and `notifyPostTasks()` of the class `PertFigure`.

We discuss calls from `PertDependency` to `PertFigures` received as parameters of type `Figure`, but deviate from our method since the decision is immediate from very general considerations. Since `start == source` and `end == target` the objects source and target are parameter objects and thus "friends", at least from the perspective of an object version of the "Law of Demeter". Since the class version of the "Law" is in general less strict[49] than the object version, it should not be more strict here. The type system forbids for good[50] but unrelated reasons to express the specific type expectations in the method declaration as the method was declared in LineConnection as handleConnect(Figure, Figure). The type expectations are clearly communicated to developers reading the method in the casts in the two first statements.

---

[49] It does not require an accessed object to actually be the calling object itself, an object held in a field, a parameter object or an instantiated object, but just to have the type of the class, any field, any parameter or any instantiated object.

I1) The essential idea here are "Downcast Parameters" since parameters can not have an arbitrary specific type. I2) Everywhere where a parameter is downcast, the type of the cast should be considered "friend". I3) For JHotDraw 5.1 it is enough to consider casts directly on parameters. I4) Given the impossibility to give covariant type information in parameter types, there is no choice but to accept the casts on parameters as type information. I5) This adaptation resolves 16 potential violations.

We could ask ourselves why we should not accept all casts. One effect would be that this could create false negatives. Such a adaptation could turn casting into a wildcard: After "(Wallet) getWallet()" it would then be acceptable to access the wallet. The smaller problem here is that one could circumvent the "Law of Demeter" checks. The bigger problem would be, that a developer might start to write code to appease some check but not to realize intended functionality.

Let us discuss all potential casts from the perspective of the "Law of Demeter". For method parameters we suggest to accept it. For constructor parameters we deny it, since the specific type could be used. For instantiated types, there is no need since the most specific type is already available. For fields we had discussed the options before with a tendency to favor keeping specialized references. Summarizing, only casts on method parameters have to be considered to be like type declarations.

**(D29) "Known Figure Parameter"**    Since the content of the Clipboard is untyped, the commands that make use of it may miss some type information as well. The first parameter of the method insertFigures(Vector, int, int) in the class FigureTransferCommand contains Figures. The same is true for the parameter of the method bounds(Enumeration k) in its subclass PasteCommand. On the elements of these enumerations the methods moveBy(int, int) and displayBox() of Figure are called. These calls from PasteCommand or FigureTransferCommand to Figures received through own methods as Objects are no violations. The type may either be inferred by analyzing the flow or we adapt the definition of the "Law" for the given methods. I1) For certain methods we know that they have "Known Figure Parameters" I2) and consider Figure a "friend" type in these specific methods I3) that we simply enumerate I4) so that we are confident in the adaption for a while. I5) This removes three potential violations.

---

[50] Let B be a subtype of A and Y a subtype of X. If Y overrides a method X.m(A) with parameter type A and the method in Y would declare the parameter type to be B, Y objects could no longer be substituted for X objects. "X x = new Y(); x.m(new A())" would be no valid code anymore. So, allowing a parameter declaration like Y.m(B) would corrupt the idea of substitutability.

■ **Listing 16** Extract from the class `FollowURLTool` in JHotDraw 5.1 (21 of 56 lines)

```
1   class FollowURLTool extends AbstractTool {
2
3       private Applet fApplet;
4
5       FollowURLTool(DrawingView view, Applet applet) {
6           super(view);
7           fApplet = applet;
8       }
9
10      public void mouseMove(MouseEvent e, int x, int y) {
11          String urlstring = null;
12          Figure figure = drawing().findFigureInside(x,y);
13          if (figure != null)
14              urlstring = (String) figure.getAttribute("URL");
15          if (urlstring != null)
16              fApplet.showStatus(urlstring);
17          else
18              fApplet.showStatus("");
19      }
20
21  }
```

The discussion in (D24) had led to the suggestion to accept accesses to `Drawing` where an access to `DrawingView` is acceptable, so that the call to findFigureInside(int, int) is considered a false positive. The given case leads to the question, whether we should decide similarly for the access to `Figure`. Allowing accesses to `Figures` where accesses to `Drawing` are allowed, would be a very general adaptation. Compared to that, the given case (a concrete tool in a sample application) is of low relevance and we suggest in (D30) to just postpone the decision temporarily and to consider the potential violation to be a false positive "for now".

### A.14  Observations in FollowURLTool

The method `mouseMove(MouseEvent, int, int)` in the class `FollowURLTool` presented in table 16 shows in the status line of an applet the URL of the figure under the mouse cursor. Since (D15) the call to findFigureInside(int, int) on the result of drawing() is considered a false positive. The subsequent call to getAttribute(String) on the result of this call is still a potential violation and will be discussed in (D30).

**(D30) Postpone Two Marginal Presumably False Positives**  JHotDraw 5.1 comes with a few sample applications that illustrate the intended usage of the framework. The application `JavaDrawViewer` uses a `FollowURLTool` that has a method `mouseMove(MouseEvent, int, int)`, presented in table 16. This method calls the method `Figure.getAttribute(String)`. Following our discussion in (D15) `DrawingView` is a "friend" of the whole class. Following our discussion in (D24) `Drawing` is a "friend" as well. We have to discuss, whether we extend this scheme so the `Figure` is a "friend" as soon as `Drawing` is.

We discuss calls from `FollowURLTool` to a `Figure` received from a `Drawing`. A1) The method `mouseMove(MouseEvent, int, int)` is coupled by the one call to getAttribute(String) to the stable interface Figure, implements the clearly understandable functionality of showing the URL of the figure at point (x, y). Since it is no secret that the Drawing contains Figures, there is no serious breach of encapsulation and the call can not be seen as a problem. A2) Since the Figure needs to be identified before calling the method lifting is not possible.[51] Pushing the call to findFigureInside(int, int) into Drawing to create a method like getFigureAttributeAt(int, int, String) would be possible. This would create one additional method, that would not be used by other classes. It would remove the two violations in this class and decouple it from Figure. Such a method fits not well to Drawing. Compared to the other methods it is too specific. It would also be technically possible to extract the last six lines of the method into a new method showAttribute(Figure, String). But the only reason for such a method would be to appease the "Law of Demeter".[52] So, neither pushing nor extracting are appropriate improvements. A3) Since the potential violation is not harmful and the possible refactorings are no improvement, the potential violation is a false positive. Unfortunately we do not see any substantial design idea behind this decision. The impact of an adaptation like "methods who are friends with Drawing are friends with Figure" would be unreasonably high compared to these two singular cases in a marginal class. Adaptations like "Figure is everybody's friend" would be even stronger and even "Figure is every tools friend" would be not justified since 10 out of 16 tools do not call a single method of Figure.

I1) We refrain from naming a design idea here since the case appears to be too singular and of low relevance. I2) To unburden the developer to revisit this unclear case regularly, we suggest to consider Figure to be a "friend" here. I3) This adaptation just extends to FollowURLTool.mouseMove(MouseEvent, int, int) and FollowURLTool.mouseDown(MouseEvent, int, int). I4) We see no reason to generalize this adaptation, but for the given methods we are very confident, that ignoring this potential violation for a while is a good solution. I5) This reduces the number of potential violations by two. The next discussion will address the last remaining potential violation.

This discussion might give an impression of how unproductive working with smell detection could feel. The case has clearly low relevance but seems to be unclear, so that the developer might need to reevaluate it again and again with no benefit. One might get tempted to apply the extraction refactoring just to appease the "Law of Demeter", but this is not a sufficient reason and turns the motivation for smell detection up-side-down. There should be no need to change the code if it does not provide any improvement.

---

[51] Lieberherr and Holland [21] named this problem as reason for the pushing refactoring.

[52] When Lieberherr, Holland, and Riel introduced the "Law of Demeter" [22], they suggested exactly this refactoring to turn any program that does not conform to the "Law" into a conforming version. They claim that this refactoring is always possible and thus any program can be transformed into an equivalent form that respects the "Law". Whether or not a program follows the "Law" would thus only be a matter of style. Unfortunately, this refactoring only somewhat improves ease of understanding, but neither encapsulation nor coupling.

■ **Listing 17**   Extract from the class ConnectionHandle in JHotDraw 5.1 (23 of 156 lines)

```
1   public class ConnectionHandle extends LocatorHandle {
2
3       private ConnectionFigure fConnection;
4       private Figure fTarget = null;
5
6       public void invokeStep (int x, int y, int anchorX, int anchorY, DrawingView view) {
7           Point p = new Point(x,y);
8           Figure f = findConnectableFigure(x, y, view.drawing());
9           // track the figure containing the mouse
10          // [ Update fTarget ]
11
12          Connector target = findConnectionTarget(p.x, p.y, view.drawing());
13          if (target != null)
14              p = Geom.center(target.displayBox());
15          fConnection.endPoint(p.x, p.y);
16      }
17
18      protected Connector findConnectionTarget(int x, int y, Drawing drawing) {
19          // [ Find a connector for a figure at location (x, y).]
20      }
21
22      private Figure findConnectableFigure(int x, int y, Drawing drawing) {
23          // [ Find figure at location (x, y).]
24      }
25
26  }
```

The call to Geom.center(Rectangle) is already because of (D4) a false positive. The method invokeStep(int, int, int, int, DrawingView) first finds a Figure at a given position and then a Connector at the same position. The former is a "friend" type because of the field fConnection, the latter not. Since there is no other comparable potential violation left and there is no obvious refactoring opportunity, we suggest in (D31) not to draw any general conclusions from this singular case but to consider it as false positive "just for now".

### A.15  Observations in ConnectionHandle.invokeStep(int, int, int, int, DrawingView)

The method invokeStep(int, int, int, int, DrawingView) in the class ConnectionHandle presented in table 17 tracks the figure under the mouse cursor and updates the end point of the connection to be created. For the calculation of the end point the method calls displayBox() on target which is of type Connector. This potential violation will be discussed in (D31).

**(D31) Postpone a Singular Possibly True Violation**    The second last line in the method invokeStep(int, int, int, int, DrawingView) in the class ConnectionHandle as presented in table 17 calls the method displayBox() of the class Connector. We discuss a call from ConnectionHandle to a Connector received from own methods accessing a Drawing. A1) This is again coupling to a rather stable interface with low intensity. The method challenges the developer to understand, why first a figure at the point (x, y) is searched but then a connector is used to specify the end point of the connection. It looks like a breach of encapsulation, when the figures share the information that they use connectors as strategy to manage the process of connecting. So, it would be nice, if understandability and encapsulation could be improved here. A2) Similar to (D30) lifting is not possible, because the connector is not uniquely defined. One would need to push at least the identification of the connector as well. `Drawing` is not a good target for this responsibility, since it is too specific and not consistent with the rest of the methods. Extracting some of the latter statements of the method into a new method with a parameter of type `Connector` is again possible but as well again too artificial. So, similar to the previous paragraph, neither pushing nor extracting are appropriate improvements. A3) The given potential violation asks for improvement, but we are not able to suggest one and we are just looking at one single isolated case. That makes postponing the decision an appropriate choice. I1) There seems to be no apparent design idea justifying the potential violation, so we just refer to this single isolated case of limited relevance. I2) We suggest to consider `Connector` to be a "friend" here for a while. I3) This adaptation just extends to `ConnectionHandle.invokeStep(int, int, int, int, DrawingView)`. I4) We see here again no reason to generalize this adaptation. Hence ignoring the potential violation for a while might give time to understand the code deeper and to decide about the case later. I5) This was the last remaining potential violation.

## About the author

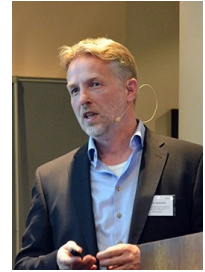

**Daniel Speicher** received his diploma in mathematics from the University of Bonn in 2003. He taught lectures in Object-Oriented Software Construction, Advanced Topics in Software Construction, and Aspect-Oriented Software Development at the Bonn-Aachen International Center for Information Technology (B-IT). He contributed to a successful series of Agile Software Development Labs as part of the International Program of Excellence (IPEC) at the B-IT. His main research interest lies in automatic code quality evaluation with reason, i.e. code quality evaluation that takes context into account, can be adapted to incorporate expert knowledge and provides explanations. Contact him at dsp@bit.uni-bonn.de.